\documentclass[a4paper,fleqn]{cas-sc}

\usepackage[authoryear]{natbib}
\usepackage{lineno}

\def\tsc#1{\csdef{#1}{\textsc{\lowercase{#1}}\xspace}}
\tsc{WGM}
\tsc{QE}
\tsc{EP}
\tsc{PMS}
\tsc{BEC}
\tsc{DE}

\begin{document}
\let\WriteBookmarks\relax
\def\floatpagepagefraction{1}
\def\textpagefraction{.001}

\shorttitle{NEAs colours}

\shortauthors{Carbognani, Buzzoni, Barbetta, Gualandi}

\title [mode = title]{Photometric characterization of near-Earth asteroids: rotation, taxonomy, and candidate evidence for internal cohesion}                      



%
\author[1]{Albino Carbognani}[orcid=0000-0002-0737-7068]

\cormark[1]


\ead{albino.carbognani@inaf.it}

\ead[url]{https://www.oas.inaf.it/}

\credit{Observations, Methodology, Computation}

\affiliation[1]{organization={INAF - Osservatorio di Astrofisica e Scienza dello Spazio},
            addressline={Via Gobetti 93/3 }, 
            city={Bologna},
            citysep={}, 
            postcode={40129}, 
            country={Italy}}

\author[1]{Alberto Buzzoni}[]



\credit{Computation}

\author[1]{Manuel Barbetta}[]



\credit{Observations}

\author[1]{Roberto Gualandi}[]



\credit{Observations}

\cortext[1]{Corresponding author}



\begin{abstract}
We present the results of a survey of previously uncharacterized near-Earth asteroids (NEAs) conducted between the second half of 2024 and the first months of 2026 using broadband $BV R_{c} I_{c}$ photometry and time-resolved lightcurve observations. Color indices were obtained for 14 NEAs of the 15 observed, with a dimension range between 0.11 and 2.5 km, and used to derive taxonomic classifications through comparison of a sample of reference reflectance curves. The sample appears to be dominated by S-, X-, and C-complex and V-type asteroids; absolute magnitudes and effective diameters were estimated using taxonomically appropriate albedos. Rotation periods were determined or constrained for several targets. Some large objects appear to rotate faster than the classical cohesionless spin barrier at $P\simeq 2.2$~h, although in some cases the period determinations remain tentative due to limited temporal coverage. A cohesive rubble-pile model provides a plausible explanation for the observed rapid rotation within the uncertainties of the derived physical parameters, and modest cohesive strengths of order $10^{2}$--$10^{4}$~Pa are sufficient to account for the observed rotation rates. 
\end{abstract}


\begin{highlights}
\item Photometry and lightcurves constrain taxonomy and rotation of near-Earth asteroids
\item Some targets rotate faster than the classical spin barrier
\item Cohesive rubble-pile model can explain the observed rapid rotation
\end{highlights}


\maketitle

\section{Introduction}
\label{sec:intro}
Over 41,000 near-Earth objects (NEOs), composed mostly of asteroids and a very small number of comets, are known in the second half of 2026, of which about 1,600 are on the risk list, i.e., a non-zero impact probability has been computed\footnote{\url{https://neo.ssa.esa.int/risk-list}}. While virtually all asteroids with a diameter of 1 km or more are known \citep{Harris2021}, the population of smaller asteroids is still largely unknown. They can hit Earth at any time, as in the cases of the Tunguska and Chelyabinsk events, caused by asteroids about 50 and 20 m in diameter, respectively \citep{Sekanina1983, Chyba1993, Borovicka2013, Popova2013, Carbognani2024}. For example, only about 44\% of asteroids with a diameter of 140 m or larger are known, and only about 10\% of asteroids with a diameter of 50 m or larger \citep{Harris2021}. From a statistical point of view, the next important collision that the Earth will suffer will be with a small-diameter asteroid, as is well exemplified by the case of 2024~YR$_4$, a NEA of about 60 m in diameter which in the second half of February 2025 reached a probability of impact with the Earth of about 3\% for 22 December 2032, now dropped to zero \citep{Bolin2025}. \\
Hence, the need is not only to discover the greatest number of NEOs to determine their orbits and compute impact probabilities, but also to characterize them physically by estimating size, mass, composition, rotation period, and structure to minimize impact risk. Observation programs, such as NEOROCKS \citep{Hromakina2021, Hromakina2023} conducted between 2020 and 2023, have addressed this need by characterizing NEAs using photometric techniques. This has led to the determination of color indices for 83 NEAs for taxonomic classification.\\
With NASA's DART mission, we have seen that using a kinetic impactor for the orbital deflection of small NEAs is possible \citep{cheng-etal_2018, daly-etal_2023}. However, the final results of the operation also depend on the structure - rubble pile with strength or monolithic - and composition of the asteroid. A direct measure of the average strength of a NEA is not easy because, for small bodies, it requires the asteroid's destruction in Earth's atmosphere, as occurred with the Chelyabinsk event. To circumvent this problem, asteroids larger than at least 150 m can be observed, thus, statistically, ruling out a monolithic structure. Given that all actual asteroids result from collisional evolution, the final structure depends on the target mass. From simulation, we know that, for small targets (150-200 m in diameter), i.e., in the strength-dominated regime, the mass of the largest intact fragment is equal to about half the target mass, and there is no reaggregation because all the fragments tend to disperse in space. Instead, for targets larger than about 300 m and regardless of material type and/or impact velocity, the mass of the largest intact fragment drops rapidly, and reaggregation occurs because collisions involving parent bodies of this size and larger occur in the gravity-dominated regime \citep{Benz1999}. This is why, statistically, small asteroids are more likely to be monolithic, while larger asteroids are generally made up of debris. This theoretical result is supported by the fact that the rotation period vs diameter shows very fast or slow rotator in the asteroid population under about 150 m in diameter, while for larger asteroids the period is generally longer than the cohesionless spin-barrier of about 2.2 h \citep{Pravec2000}, the limit below which a rubble pile without internal cohesion can be disrupted. Rotation periods under this critical value will cause asteroid breakup and the formation of a binary system \citep{Pravec2007}, or an asteroid pair \citep{Carbognani2026}. \\
So, measuring the rotation period and using a model of an asteroid with a cohesive rubble-pile structure \citep{Sanchez2014}, a lower-order estimate of the asteroid’s strength can be inferred, subject to assumptions on shape, density, and internal structure. These data can be used statistically to better predict the effects of orbital deflection techniques that can be applied to minimize the risk of collision with Earth. Knowing an asteroid's rotation period can also be important for planning a mission to its surface: if rotations are too rapid, centrifugal acceleration can exceed gravitational acceleration, making it difficult for a lander to descend to the surface, as in the case of 2022~OB$_5$ \citep{Alarcon2026}.\\
This paper presents our results about the physical characterization of some NEAs observed between 2024 and 2026 with multiband $BVR_cI_c$ photometry. The NEAs to be observed are uncharacterized or partly characterized, and were chosen based on the brightness and favorable observation conditions from the observing site, with the main goal of identifying NEAs with rotation periods shorter than the cohesionless spin barrier, in order to estimate the minimum strength value, and the taxonomic classes.  \\
The paper is organized as follows: in Section~\ref{sec:obs}, we will see the instruments used; in Section~\ref{sec:met}, we illustrate the method for the taxonomic classification of the asteroids and the dimensions estimate, while in Section~\ref{sec:dis}, the main results obtained are exposed. To avoid making the reading difficult, we have included the observational circumstances and the results for each observed asteroid in Appendix~\ref{sec:appendix}. Finally, in Section~\ref{sec:end}, we will provide our conclusions.

\section{Instruments}
\label{sec:obs}
For our observations, we used mainly two different telescopes. The largest instrument was the ``G.D. Cassini'' 1.52-m F/4.8 Ritchey-Chr\'etien telescope of the Loiano Astronomical Station (IAU 598), managed by the Astrophysics and Space Science Observatory of Bologna (Italy). The Bologna Faint Object and Spectroscopic Camera (BFOSC) was attached to the telescope, equipped with a Princeton Instruments EEV $1340 \times 1300$ pixel back-illuminated CCD with 20~$\mu$m pixel size. The plate scale in bin one mode was 0.58~arcsec~px$^{-1}$ leading to a field of view of $13.0 \times 12.6$~arcmin. Broad-band Johnson/Cousins $BVR_{c}I_{c}$ filters were used to measure the asteroid's colors. This instrument, which has a magnitude limit of +22.5 with about twenty minutes of exposure without a filter, was also used recently to characterize the active main-belt asteroid (6478) Gault \citep{Carbognani2021}. Unfortunately, it was not always available to observe NEAs due to other scientific programs.\\
The second instrument was TANDEM, the Telescope Array eNabling DEbris Monitoring (IAU D98). TANDEM consists of a combo of four customized and independently steerable 35 cm f/3 Newtonian telescopes, each equipped with a Moravian C4-16000 camera, observing through the $BVR_{c}I_{c}$ filters of the Johnson-Cousins system \citep{Buzzoni2025}. The camera is equipped with a GSense 4040 ($4096\times 4096$ px) monochrome CMOS sensor, featuring an electronic shutter and a pixel size of $9 \mu\text{m}$. A corrected field of view (FOV) of $2^\circ \times 2^\circ$ is offered by each telescope, though quite special pointing capabilities and observing modes are available for the telescope array, such as to cover up to $16$ square degrees across sparse celestial fields. While specifically designed for observing activities within the European Consortium for Space Surveillance and Tracking (EU-SST) framework \citep{Peldszus2022}, TANDEM may also have additional applications in a more direct astronomical context, as the astrometry of very fast and bright NEAs \citep{Palmiotto2026}. The great advantage of TANDEM over a single-tube telescope is that it can image the target simultaneously in four colors. In this way, it becomes possible to study the potential variation in the color indices as a function of the rotational phase without changing the filter, as in a single telescope. Unfortunately, it is not possible to use ``Cassini'' simultaneously with TANDEM, and the small diameter of a single telescope in the TANDEM system limits observations to only bright targets.\\
In one case, the 0.60-m F/4.0 Newtonian telescope of the San Marcello Pistoiese Observatory (IAU 104) was used to determine the rotation period of the NEA (488693) 2003 WW$_{87}$. This telescope has an Apogee $1024 \times 1024$ pixel CCD camera with 24~$\mu$m pixel size. In unbinned mode, the detector provided a plate scale of 2.1~arcsec~px$^{-1}$ and a field of view of $35.2 \times 35.2$~arcmin with a magnitude limit of about +21 in twenty minutes exposure. Also, the 0.30-m F/4.0 Newtonian telescope of the Virgil Observatory (IAU M60) near Loiano Astronomical Station was sometimes used to complete the lightcurves obtained with ``Cassini''. This small telescope is equipped with an ASI 533MM CMOS camera (plate scale of 1.3~arcsec~px$^{-1}$ in bin 2 mode), and a $BVR_{c}C$ filter wheel. Due to the severe light pollution problem which affects the small village of Loiano, this telescope can reach the magnitude +18.5 in a twenty-minute exposure.\\

\section{Methods}
\label{sec:met}
Regarding the $BVR_{c}I_{c}$ color measurements, the images were all acquired during nights with stable atmospheric transparency. Photometric reduction was performed following standard calibration procedures, including bias subtraction, dark correction, and flat-fielding. Photometric calibration was carried out using Landolt standard fields \citep{Landolt1992, Harris1981}. Special care was taken to observe the Landolt fields with airmasses similar to those of the asteroid frames, to minimize differential airmass effects. Initially, the image processing required to compute color indices was performed using the ESO--MIDAS software package \citep{banse83}. In the photometric observations, alternating sequences of $BVR_{c}I_{c}$ filters were taken at least two or three times, so that the apparent magnitude of the asteroid in different colors could be determined at the same common time using a linear least-squares fit. This compensates for the asteroid's change in magnitude due to spin. \\
Later, in addition to the ESO--MIDAS software package, a dedicated and more efficient Python script was developed to calibrate the following linear transformation using Landolt standard stars at the same asteroid's airmass \citep{Harris1981, Carbognani2019}:

\begin{equation}
m_{\mathrm{std}}(X) - m_{\mathrm{inst}}(X) = ZP(X) + CT(X)\cdot CI,
\label{eq:calib}
\end{equation}

\noindent In Eq.~\ref{eq:calib} $m_{\mathrm{std}}(X)$ is the standard magnitude of a Landolt star in the filter $X = B, V, R_{c}, I_{c}$, $m_{\mathrm{inst}}(X)$ is the corresponding measured instrumental
magnitude, $ZP(X)$ is the photometric zero point that also incorporates the fixed airmass dependence, and $CT(X)$ is the colour term. The color index $CI$ is defined as $(B - V)$ for $X = B$ and $X = V$, $(V - R_c)$ for $X = R_{c}$, and $(R_c - I_c)$ for $X = I_{c}$. Once the coefficients $ZP(X)$ and $CT(X)$ were determined for each filter by solving a linear system, they were used to compute the calibrated asteroid magnitudes at the same airmass as Landolt's field. Since the asteroid color indices are initially unknown, Eq.~\ref{eq:calib} was iterated a few times until convergence, using color indices derived from the instrumental magnitudes as initial values. \\
We verified that the asteroids' color indices obtained through this last approach are consistent, within the measurement uncertainties, with those derived from the ESO--MIDAS software package. The consistency was verified for the NEAs 2003~MX$_2$, 2003~WW$_{87}$, 2011~UL$_{21}$, and 2024~WB. We have also verified that the $B-V$ values for the main-belt asteroids (244) Sita, (784) Pickeringia, (61) Danae, and (322) Phaeo given by the JPL Small-Body Database\footnote{\url{https://ssd.jpl.nasa.gov/tools/sbdb_lookup.html}}, are equal to that find by us within 1-3 hundredths of a magnitude. 

\subsection{Rotation period determination}
\label{sub:period}
As regards the determination of the rotation period of the asteroids, dense photometric sessions were carried out with the $R_c$ filter. The photometric observations were reduced using the MPO {\sc Canopus} software package\footnote{\url{http://minorplanetobserver.com}}, taking advantage of its special implementation of the classical FALC algorithm \citep{Harris1989} for period determination starting from the observed sessions' lightcurves.\\
For asteroids observed in a single session, we analyzed the reduced photometric data obtained with {\sc Canopus} using a home-made Python software for period search based on the Lomb--Scargle (LS) algorithm, widely used for detecting and characterizing periodicity in unevenly sampled time-series \citep{Lomb1976, Scargle1982, VanderPlas2018}. The LS periodogram was computed using a multi-term Fourier model with 2 harmonics (in order to better fit the non-sinusoidal lightcurve), over a range of trial periods, and the strongest peak corresponding to a preliminary period $p$ was identified. The false-alarm probability (FAP) for the best period was computed using the bootstrap method, which produces the most robust estimate because it makes few assumptions about the form of the periodogram. The FAP value of a period $p$ in the periodogram is useful because it tells us the probability that a lightcurve with no signal gives a peak in the periodogram as high as the found period $p$ \citep{VanderPlas2018}. Generally, if the FAP is 0.01 or lower, the period can be considered real. A weighted least-squares fit with a Fourier series of order $n$ (typically $n = 2$ as LS) was then performed for the candidate periods $P = p/2$, $P = p$, and $P = 2p$. The period corresponding to the minimum reduced $\chi^2$ value of the Fourier fit was adopted as the best estimate of the rotation period: usually, the best period from LS and the $P$ value from minimum $\chi^2$ are the same, as expected. From the best-fitting Fourier model, the peak-to-peak amplitude of the lightcurve was derived. \\
In general, the rotation periods obtained with the FALC algorithm and those derived using the LS-based analysis described here are mutually consistent. In cases where the phased lightcurve exhibited significant dispersion above the point uncertainties, the LS analysis was also used as a phenomenological tool to explore the presence of a possible second periodicity. For this purpose, the lightcurve was modeled as the sum of two independent Fourier series corresponding to two trial periods, and the couple with the minimum $\chi^2$ was selected. This approach is intended as a diagnostic method to identify complex rotational behavior and does not represent a full physical model of non--principal--axis rotation, which would require the inclusion of cross-terms between the two rotation frequencies (e.g., \citealt{Pravec_2005}).\\
Finally, we assign a lightcurve quality code ($U$) following the convention of \cite{Warner2009}, adopted in the Asteroid Light Curve Database (ALCDB), based on phase coverage, uniqueness of the period solution, and consistency across datasets. Values range from $U = 0$ (no reliable period) to $U = 3$ (secure solution).

\subsection{Taxonomic classification}
\label{sub:tax}
One of the main challenges with multi-color photometric observations of asteroids is determining their taxonomic class. For this purpose, we computed the reflectance values and compared them with the reflectance curves of asteroids of known taxonomy. The reflectance values are given by the following equation \citep{DeMeo2013}:

\begin{equation}
    R_f=10^{-0.4\left[\left(f-V\right)-\left(f-V\right)_\odot\right]}
     \label{eq:reflect}
\end{equation}

\begin{figure}
    \centering
    \includegraphics[width=1.0\textwidth]{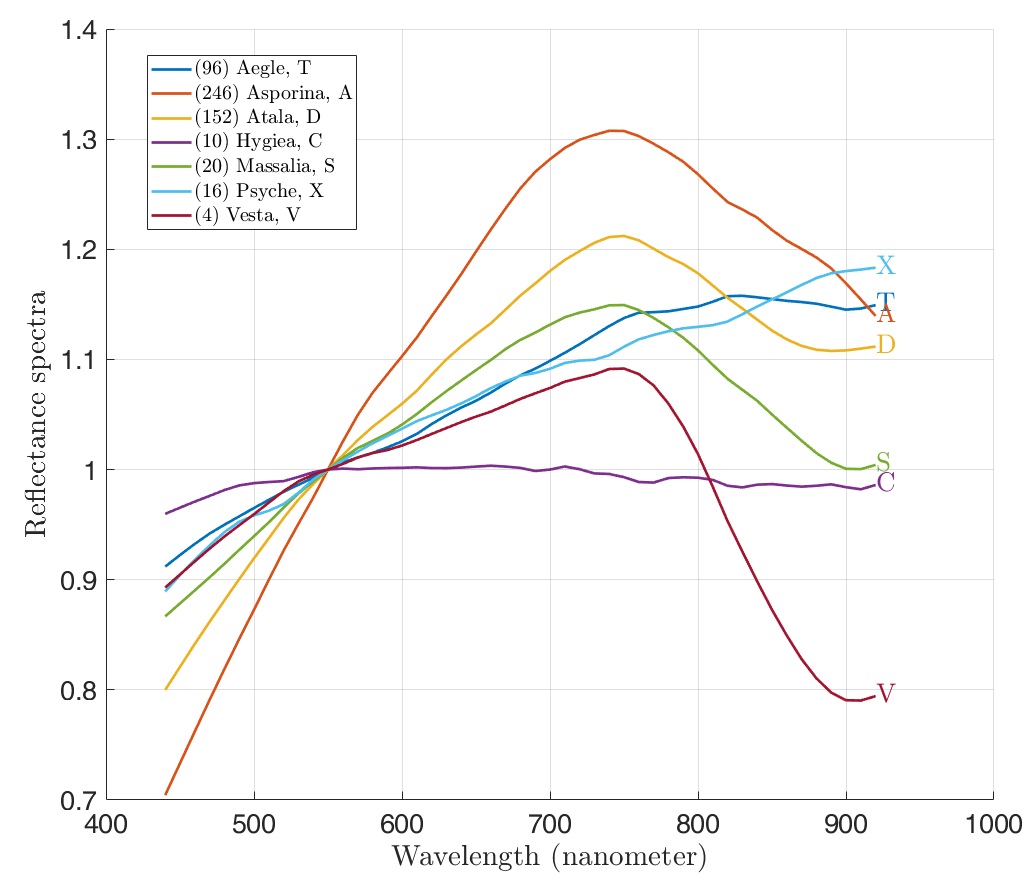}
    \caption{Some reflectance spectra from SMASS II for the asteroid taxonomic classes considered in the text. While the slopes between 750 and 450 nm are quite similar, excluding the C-complex, the curves are well differentiated towards the near-infrared, which facilitates asteroid classification when the magnitude in this band is also measured. All the reflectance spectra are normalized to one at 550 nm.}
    \label{fig:reflectance}
\end{figure}

\noindent In Eq.~\ref{eq:reflect} $f$ and $f_\odot$ are the magnitude of the asteroid and the Sun in a fixed filter. The equation is normalized to unity at the wavelength of 550 nm. The adopted colour indices of the Sun are $\left(B-V\right)_\odot=0.642$, 
$\left(R-V\right)_\odot=-0.354$ and $\left(I-V\right)_\odot=-0.688$ \citep{Holmberg2006}. \\
As a reference, we adopted 1330 reflectance curves\footnote{\url{http://smass.mit.edu/data/smass/smass2/}} observed during the second phase of the Small Mainbelt Asteroid Spectroscopic Survey (SMASS II). These reflectances, with visible wavelength spectroscopy between 450 and 950 nm, formed the base for the Bus and Binzel asteroid taxonomy \citep{Bus2002}. This classification comprises 26 classes, but since we use broadband photometry, we consider only the main taxonomic classes: S-complex (including L-type), C-complex (including B-type), X-complex, V-type, D-type, and T-type. In our reference sample, the S-complex is about 41\% of the total, the C-complex is 25\% of the total, and the X-complex is 18\%, with the remaining classes that make up the other 16\%. There are small variations in the reflectance curve within the same class, but in general they are well defined, see \cite{DeMeo2013}. The reflectance curves show a slightly different slope between 750 and 450 nm, except for the C-complex, which has a substantially flat curve, see Fig.~\ref{fig:reflectance}. The A, V, S, and D reflectance curves rise to varying degrees between 900 and 750 nm, while the X and T curves' slopes remain stationary, and the two types are difficult to separate. This different reflectance pattern can provide an indication of the taxonomic class, although degeneracies remain when only broadband photometry is available.\\
For each asteroid, we adopted the classification that minimizes the distance between the observed four-point reflectance given by Eq.~\ref{eq:reflect} and the full reflectance curve:

\begin{equation}
    d=\sqrt{\left(R_B-Re(B)\right)^2+\left(R_V-Re(V)\right)^2+\left(R_R-Re(R)\right)^2+\left(R_I-Re(I)\right)^2 }
    \label{eq:col}
\end{equation}

\noindent In Eq.~\ref{eq:col}, $R_f$ with $f=B, V, R_c, I_c$ are the observed reflectance at the filter effective wavelength\footnote{\url{https://www.aavso.org/filters}}, respectively, of 436.1, 544.8, 640.7, and 798 nm, while $Re(f)$ is the reference reflectance curve at the same wavelength. Note that this approach provides a best-fit solution rather than a unique classification. \\
We estimated the robustness of each taxonomic assignment through a Monte Carlo procedure. For each asteroid, 500 synthetic color sets were generated by sampling the measured color indices within their Gaussian uncertainties. Each color set was converted to reflectance and classified using the same minimum-distance criterion adopted for the nominal solution. The taxonomic probability was then defined as the fraction of realizations assigned to each class. The robustness percentage of the adopted classification is reported in brackets in Table~\ref{tab:Neas}. The probabilities are nearly equal even with only 100 synthetic color sets, indicating that the Monte Carlo sampling is adequate. Note that these probabilities represent relative classification likelihoods within the adopted template set and do not constitute absolute taxonomic probabilities.

\subsubsection{The Phase Reddening Effects}
\label{sub:tax_red}
A further challenge in deriving taxonomic classifications of asteroids is the influence of viewing geometry on the observed reflectance slope, a phenomenon known as phase reddening. As the phase angle ($\alpha$) increases, multiple scattering processes within the surface regolith alter the reflectance, typically causing an asteroid to appear systematically redder than at opposition, i.e., when $\alpha = 0^\circ$. The phase reddening effect is most pronounced at wavelengths longer than $1~\mu\text{m}$ and is strongly coupled to the target's surface composition, mineralogy, and albedo. Silicate-rich bodies, such as those belonging to the X-complex and V-type, exhibit well-documented, steep increases in their color indices with phase angle. Conversely, primitive, low-albedo carbonaceous objects (C-complex) display nearly negligible variations within the visible regime \citep{Sanchez2012, Perna2018}. For example, referring to Fig.~\ref{fig:reflectance}, we can see that, if you increase the slope a little towards the IR, a V-type can change into an S-complex, or an X-complex can become a D-type. \\
The phase reddening cannot be negligible for NEAs, as they can be observed at large phase angles, so addressing this effect is critical to identify geometric biases that could lead to misclassifications if we determine the taxonomic class only from one position, as in our case. The best way to discriminate the phase reddening effect is to determine the taxonomic class as a function of phase angle, but this is not always possible because it requires substantial telescope time. \\
In this section, starting with the reddening of the reflectance slope, we estimate taxon-specific phase reddening coefficients ($\beta$) for color indices that can correct high-phase observations to a standard opposition geometry when the asteroid's taxonomy is known in advance. Of course, this is not possible if the asteroid is observed for the first time; however, these coefficients remain useful because they indicate which taxonomies are more strongly influenced by the reddening-phase effect in the color-index regime.\\
So, we want to convert the spectroscopic phase-reddening coefficients ($\beta_{\text{slope}}$) reported for asteroids' spectra in the wavelength range $0.44-0.65~\mu\text{m}$ by \cite{Perna2018} into directly applicable photometric variations ($\beta_{\text{color}}$). Starting from Eq.~\ref{eq:reflect}, a generic photometric color index $(f_1 - f_2)$ is related to the ratio of the asteroid reflectances in the central wavelengths of the two filters ($\lambda_1$ and $\lambda_2$) by:

\begin{equation}
(f_1 - f_2) = (f_1 - f_2)_{\odot} - 2.5 \log_{10} \left[ \frac{R_{f1}(\lambda_1)}{R_{f2}(\lambda_2)} \right]
\label{eq:color_reflectance}
\end{equation}

\noindent If we assume that the reflectance spectra of our targets are mostly approximately linear in the optical range, we can approximate the reflectance $R_{f}(\lambda_1)$ normalized at 1 to reference $\lambda_0$ with:

\begin{equation}
R_{f}(\lambda_1)\approx 1+S(\alpha)(\lambda-\lambda_0)\approx 1+(S_0+\beta_{\text{slope}}\alpha)(\lambda-\lambda_0)
\label{eq:reflectance}
\end{equation}

\noindent In the Eq.~\ref{eq:reflectance}, $S(\alpha)$ is the linearized spectrum slope. Substituting Eq.~\ref{eq:reflectance} in Eq.~\ref{eq:color_reflectance}, and deriving it with respect to $\alpha$, we have a first-order analytical relation linking broadband colors correction to spectral slopes:

\begin{equation}
\beta_{\text{color}} = \frac{d(f_1-f_2)}{d\alpha} \approx -1.1 \beta_{\text{slope}} (\lambda_1 - \lambda_2)
\label{eq:color_slope}
\end{equation}

\noindent In Eq.~\ref{eq:color_slope}, $\lambda_1$ and $\lambda_2$ are the effective wavelengths of the filters in micrometers. If we know $\beta_{\text{color}}$, we can correct the observed color index $CI(\alpha)$ to $\alpha=0$ using the linear approximation $CI(0)\approx CI(\alpha)-\beta_{\text{color}}\alpha$. In Table~\ref{tab:slope_color}, using the reflectance slope, we have estimated the $\beta_{\text{color}}$ values for our color indices. From this Table, considering the mean value of the $\beta_{\text{color}}$ reported in the penultimate column, we can see that for an S-, C-complex and D-type asteroid, the color variation is practically negligible at $\alpha=90^\circ$, with a change of about a few hundredths of a mag, a value similar to the color uncertainty. Instead, for a Q-type and A-type asteroid, the change in color is about 0.3-0.8 mag, a value higher than the typical mag uncertainty of color indices. The same thing holds for the V-type and X-complex, where the change in the same condition is about 0.1 mag. So these are the four asteroid taxonomies that can change due to the reddening effect for high phase angle.

\begin{table*}
	\centering
	\caption{Approximate conversion from $\beta_{\text{slope}}~(\mu\text{m}^{-1} \text{deg}^{-1})$ from \cite{Perna2018}, measured in the wavelength range $0.44-0.65~\mu\text{m}$, to $\beta_{\text{color}}~(\text{mag/deg})$ for different taxonomy and color index using the linear approximation discussed in the text (see Eq.~\ref{eq:color_slope}). The data for the V-type asteroid, instead, came directly from a best-fit linear regression of the color indices taken with $BVR_cI_c$ filters; see \cite{Hicks2014}. In the last two columns, there is the mean $\bar{\beta}_{\text{color}}$ and the estimated mean variation of the color indices for $\alpha=90^\circ$. The A-type and Q-type asteroids are more subject to the reddening effect, followed by the V-type and X-complex.}
	\label{tab:slope_color}
        \setlength\tabcolsep{2pt} 
	\begin{tabular}{lcccccccc} 
\hline
Tax & $\alpha$ range (deg) &	$\beta_{\text{slope}}$	& $\beta_{B-R_c}$ & $\beta_{B-V}$ & $\beta_{V-R_c}$  & $\beta_{V-I_c}$   & $\bar{\beta}_{\text{color}}$ & $CI(0^\circ)-CI(90^\circ)~(\text{mag})$\\
\hline
C & 02-69 & -0.0032 & -0.00071 & -0.00038 & -0.00033 & -0.00088  & -0.00057 & -0.05\\
S & 04-92 & 0.0013  & 0.00029  & 0.00015  & 0.00013  & 0.00036   & 0.00023  & 0.02 \\  
X & 05-56 & 0.0074  & 0.0016   & 0.00087  & 0.00077  & 0.0020    & 0.0013   & 0.12 \\
V & 26-62 &  ---    & 0.0023   &   ---    & 0.00075  &   ---     & 0.0015   & 0.14\\
A & 20-51 & 0.0492  & 0.011    & 0.0058   & 0.0051   & 0.013     & 0.0087   & 0.78\\
D & 06-64 & -0.0024 & -0.00053 & -0.00028 & -0.00025 & -0.00066  & -0.00043 & -0.04\\
Q & 10-80 & 0.0074  & 0.0016   & 0.0087   & 0.00077  & 0.0020    & 0.0033   & 0.30\\
\hline
	\end{tabular}
\end{table*}

\subsection{Dimension and shape estimate}
\label{sub:dimension_estimate}
Once the taxonomic class of the asteroid has been determined, the mean geometric albedo values of the class can be used to estimate its actual size. However, to estimate the size, the geometric albedo is not enough; it is also necessary to know the absolute magnitude.\\
For the estimate of $H_V$, in first approximation, we used the mean value of the magnitude in $V$ using the simple $H-G$ phase-mag curve defined by \cite{Bowell1989}:

\begin{equation}
    H_V=m_V(\alpha)+2.5log_{10}\left((1-G)\phi_1(\alpha)+G\phi_2(\alpha)\right)
    \label{eq:HG}
\end{equation}

\noindent In Eq.~\ref{eq:HG}, $m_V(\alpha)$ is the reduced $V$ magnitude from observations, $\phi_i(\alpha)$ with $i=1, 2$ are known functions of the phase angle $\alpha$ and the mean $G$ values for a taxonomic class is given by \cite{Veres2015}, see  Table~\ref{tab:Tholen}. In an advanced step, we used a model phase-mag curves more sophisticated, the ($H$, $G_1$, $G_2$) system by \cite{Muinonen2010}:

\begin{equation}
    H_V=m_V(\alpha)+2.5log_{10}\left(G_11\Phi_1(\alpha)+G_2\Phi_2(\alpha)+(1-G_1-G_2)\Phi_3\right)
    \label{eq:HG1G2}
\end{equation}

\noindent In Eq.~\ref{eq:HG1G2} $\Phi_i(\alpha)$ with $i=1, 2, 3$ are known functions of the phase angle, of which we have taken the tabulated values from \cite{Penttila2016}. The main advantage of the $H$, $G_1$, $G_2$ system over the classical $HG$ system is that it better describes the true shape of asteroid phase curves, especially at small phase angles near the opposition effect. For $G_1$ and $G_2$, we use the most recent taxonomic-class values from \cite{Colazo2025}. These values were obtained from 301~272 asteroids in the orange filter and 280~953 in the cyan filter present in the ATLAS Solar System Catalog V2 (SSCAT-2). Considering that the wavelength range of the orange filter is 560-820 nm, while the cyan range is 420-650 nm, we took the $G_1$ and $G_2$ values from the phase-mag curve for the last color, the one closest to the $V$ filter than the previous one. The paper of \cite{Colazo2025} explicitly says that Table 2, which show the values of $G_1$ and $G_2$, reports the geometric center of the 95\% probability contour and the area of the $1-\sigma$ contour for each taxonomic class/filter, so we have estimated only a global 2D uncertainty region, not independent uncertainties on $G_1$ and $G_2$, assuming the $1-\sigma$ region is circular with $\delta G_{1,2}=\sqrt(area/\pi)$. This is only a rough isotropic approximation, because the actual contours from \cite{Colazo2025} are elongated, so the real uncertainty is anisotropic. The adopted $G_1$ and $G_2$ wit estimated uncertainty values are given in Table~\ref{tab:Tholen}.\\  
Once the absolute magnitude $H_V$ has been estimated, using the mean geometric albedo $p_V$ corresponding to the taxonomic class \citep{Usui2013}, the diameter estimate in km can be obtained with the equation \citep{Harris2002}:

\begin{equation}
    D=\frac{1329}{\sqrt{p_V}} 10^{-H_V/5}
    \label{eq:diameter}
\end{equation}

\noindent bearing in mind that the use of class-averaged albedos introduces significant uncertainties on individual objects, which we have taken into account in the propagation of errors. We will explicitly mention the use of these equations for the first two asteroids described in the appendix; for the following ones, their use will be implied. It should be emphasized that using the geometric albedo for a given taxonomic class does not guarantee that the asteroid actually has that geometric albedo. There may be variations within a class \citep{Masiero2011}, but a diameter estimate, even an approximate one, is always better than no estimate at all, especially for near-Earth asteroids.\\

\begin{table*}
	\centering
	\caption{Summary of the adopted mean geometric albedo $p_V$, $G$, $G_1$, and $G_2$ for the most common taxonomic classes used to estimate the dimension. The values of $p_V$ came from \cite{Usui2013}; $G$ values from \cite{Veres2015}; $G$ value for V-type is from \cite{Oszkiewicz2021}, while The T values are from \cite{Warner2009}. The $G_1$ and $G_2$ values adopted for the $HG_1G_2$ model were taken from \cite{Colazo2025}, except for T, which was taken from \cite{Mahlke2021}.}
	\label{tab:Tholen}
        \setlength\tabcolsep{2pt} 
	\begin{tabular}{cccccc} 
\hline
Taxonomic class &	$p_V$	& $G$    &    & $G_1$   & $G_2$	 \\
\hline
C & $0.071 \pm 0.04$  & $0.15 \pm 0.09$  & & $0.40 \pm 0.18$    & $0.36 \pm 0.18$  \\
S & $0.208 \pm 0.079$ & $0.24 \pm 0.06$  & & $0.26 \pm 0.14$    & $0.50 \pm 0.14$  \\
X & $0.098 \pm 0.081$ & $0.20 \pm 0.09$  & & $0.38 \pm 0.16$    & $0.38 \pm 0.16$  \\  
V & $0.297 \pm 0.131$ & $0.35 \pm 0.09$  & & $0.22 \pm 0.11$    & $0.56 \pm 0.11$  \\
D & $0.086 \pm 0.053$ & $0.09 \pm 0.09$  & & $0.37 \pm 0.16$    & $0.42 \pm 0.16$  \\
T & $0.057 \pm 0.02$  & $0.12 \pm 0.08$  & & $0.65 \pm 0.14$    & $0.19 \pm 0.14$  \\
\hline
	\end{tabular}
\end{table*}

When the lightcurve amplitude is determined, we have estimated the lower limit of the ratio between axes of the asteroid, which is supposed to be an ellipsoid shape. However, we cannot take the amplitude from the observed lightcurve directly because we must first correct it. A so-called amplitude-phase relationship (APR) is recognized for asteroids, and for phase angle $\alpha \lesssim 40^\circ$ the relation is linear and of the form \citep{zappala1990}: 

\begin{equation}
A(0) = \frac{A(\alpha)}{\left(1+ m\alpha\right)}
\label{eq:zap}
\end{equation}

\noindent In the Eq.~\ref{eq:zap}, $A(0)$ is the lightcurve amplitude (in mag) at the opposition (namely at $\alpha = 0^\circ$), while $A(\alpha)$ is the observed amplitude that increases with the phase angle starting from the opposition value. If we express $\alpha$ in degrees, then the scaling coefficient $m$ can be empirically calibrated. According to \cite{zappala1990}, the $m$ value depends on the taxonomic type, with $m = 0.030, 0.015$, and 0.013, respectively, for S-, C-, and M-type asteroids. More recent studies instead indicate the asteroid's obliquity\footnote{The obliquity is the angle between a plane X, containing the observer and the spin vector, and the normal to a plane Y containing the asteroid, the observer, and the Sun.} as the most important parameter from which the $m$ value depends, with a negligible contribution from surface texture and taxonomic type \citep{Gutierrez2006}. With an average obliquity $\gamma=60^\circ$, expected for an asteroid population with fully randomized rotational axes, the $m$ value is estimated to be comprised between 0.010 and 0.018 $\text{mag deg}^{-1}$. As an operative value, we take the mean, i.e. $m=0.014 \text{mag deg}^{-1}$, to estimate $A(0)$ from Eq.~\ref{eq:zap}, independently of the taxonomy. From this opposition amplitude, the true inferior limit of the ratio of asteroid axis can be estimated by $a/b \geq 10^{0.4A(0)}$. Of course, this value must be taken with caution, as we do not know the true obliquity angle; it is valid only in a statistical sense.

\begin{table*}
	\centering
	\caption{Summary of the near-Earth asteroids observed with BFOSC (B) and TANDEM (T) with mean $V$ mag at the time of observation, mean color indices, and the taxonomic class derived from the observed reflectance curve, comparing it with the reflectance curves database. The percentage between parentheses is the maximum probability of the taxonomic classification obtained from a Monte Carlo computation, as explained in the text. An asterisk near the classification indicates an uncertainty; to find out the specific reasons, please refer to the comments dedicated to each asteroid in the appendix.}
	\label{tab:Neas}
        \setlength\tabcolsep{2pt} 
	\begin{tabular}{lccccccccccc} 
\hline
NEA &	Obs Date	& r, au 	& $\Delta$, au & $\alpha$ & Inst. & $\overline{V}$ & $B-R_c$ & $B-V$ & $V-R_c$ & $V-I_c$ & Taxonomy \\
\hline
2011~UL$_{21}$ & 30 Jun 2024 & 1.054 & 0.063 & 51.8 & T & 12.1  & $1.49 \pm 0.01$ & $0.93\pm 0.01$ & $0.56\pm 0.01$ & $0.88\pm 0.01$ & S (100\%)\\
          & 12 Jul 2024 & 1.175 & 0.229 & 42.0 & B  & 14.8  & $1.32 \pm 0.01$ & $0.85\pm 0.01$ & $0.47\pm 0.01$ & $0.78\pm 0.01$ & S (49\%)\\
2024~MK   & 30 Jun 2024 & 1.016 & 0.008 & 95.1 & T & 15.0  & $1.36 \pm 0.02$ & $0.91\pm 0.02$ & $0.45\pm 0.02$ & $0.80\pm 0.02$ & S (67\%)\\
2015~RG$_{36}$ & 11 Nov 2024 & 1.085 & 0.100 & 16.8 & B  & 16.5  & $1.34 \pm 0.1$  & $0.81 \pm 0.06$& $0.53 \pm 0.1$ & $0.9 \pm 0.1 $ & S (60\%)\\
2003~WW$_{87}$ & 15 Nov 2024 & 1.124 & 0.204 & 44.8 & B  & 15.2   & $1.21 \pm 0.05$ & $0.77 \pm 0.05$& $0.43 \pm 0.02$& $0.76\pm0.03$  & S (42\%)\\
2003~MX$_2$  & 02 Dec 2024 & 1.480 & 0.540 & 19.1 & B  & 17.0   & $1.31\pm 0.07$  & $0.89\pm 0.09$ & $0.41\pm 0.02$ & $0.84\pm 0.1$  & S (50\%)\\
2024~WB   & 02 Dec 2024 & 1.068 & 0.082 & 06.0 & B  & 16.2   & $1.15 \pm 0.1$  & $0.83 \pm 0.1$ & $0.32 \pm 0.04$& $0.56 \pm 0.04$& C (83\%)\\
2021~PS$_2$  & 07 Mar 2025 & 1.099 & 0.120 & 24.2 & B  & 16.4   & $1.10 \pm 0.1$ & $0.82 \pm 0.1$ & $0.27 \pm 0.01$ & $0.53 \pm 0.08$ & C (88\%)\\
(9058) 1992~JB & 30 Apr 2025 & 1.107 & 0.120 & 32.0 & B & 15.0 & $1.30 \pm 0.02$ & $0.84 \pm 0.02$ & $0.46 \pm 0.01$ & $0.85 \pm 0.02$ & S (77\%)\\
               & 26 May 2025 & 1.229 & 0.281 & 35.5   & B      &        17.0         &                &                &                &    & \\
2005~VO$_5$ &27 Jun 2025  & 1.134 & 0.120 & 12.0   & B      &        16.1         &       ---       &         ---     &      ---       &   ---  & ---\\
(333284) 1999~PJ$_1$  & 22 Aug 2025 & 1.323 & 0.333 & 18.2 & B  & 17.3   &  $1.63 \pm 0.15$ & $1.19 \pm 0.09$ & $0.44 \pm 0.08$ & $0.60 \pm 0.06$ &   V* (81\%)\\
(152664) 1998~FW$_4$  & 18 Sep 2025 & 1.125 & 0.123 & 11.0 & B  & 15.9   & $1.19 \pm 0.06$ & $0.75 \pm 0.03$ & $0.44 \pm 0.03$ & $0.77 \pm 0.14$ & X (14\%)\\
        2025~FA$_{22}$  & 26 Sep 2025 & 1.050 & 0.053 & 25.3 & B  & 16.2  & $1.13 \pm 0.04$ & $0.66 \pm 0.05$ & $0.46 \pm 0.03$ & $0.86 \pm 0.04$ &   X (70\%)\\
(138205) 2000~EZ$_{148}$  & 16 Oct 2025 & 1.453 & 0.479 & 14.9 & B & 15.3  &  $1.34 \pm 0.04$ & $0.84 \pm 0.03$ & $0.503 \pm 0.003$ & $0.91 \pm 0.02$ & S (98\%)\\
(843079) 2016~CB$_{32}$  & 29 Nov 2025 & 1.183 & 0.202 & 12.9 & B & 16.4  &  $1.18 \pm 0.05$ & $0.72 \pm 0.03$ & $0.46 \pm 0.02$ & $0.77 \pm 0.02$ & X (44\%)\\
(52340) 1992~SY  & 17 Feb 2026 & 1.026 & 0.134 & 70.0 & B & 15.9  &  $1.29 \pm 0.08$ & $0.87 \pm 0.08$ & $0.43 \pm 0.02$ & $0.71 \pm 0.04$ & V* (45\%)\\
        \hline
	\end{tabular}
\end{table*}

\begin{table*}
	\centering
	\caption{Summary of the near-Earth asteroids rotation period, the $U$ quality code, amplitude, mean absolute mag $H_V$, the effective diameter $D$ (both from $HG_1G_2$ model), and the minimum axes ratio $a/b$. In the column marked with ``Note'', we report whether the rotation period was known or unknown at the time of observations. If already known, the period values are compared with the Small-Body Database (SBDB) of the NASA Jet Propulsion Laboratory (JPL), which collects information from the ALCDB and peer-reviewed publications. The letter B or T next to the asteroid name stands for BFOSC or TANDEM. The presence of a question mark indicates an uncertain outcome for the rotation period and an application of the amplitude-phase relationship beyond the phase-angle limit.}
	\label{tab:Neas_period}
        \setlength\tabcolsep{2pt} 
	\begin{tabular}{lccccccl} 
\hline
NEA &	Rotation Period (h)	& $U$ & A (mag) 	& $H_V$ (mag) & $D$ (km)& $a/b \geq$ & Note  \\
\hline
(415029) 2011~UL$_{21}$ (T) & - & - & - &  $16.4 \pm 0.3$ & $1.5 \pm 0.3 $    & - & Period of about 2.5 h (radar)\\
(415029) 2011~UL$_{21}$ (B) & - & - & - &  $16.3 \pm 0.3$ & $1.6 \pm 0.4 $    & - &                               \\
         2024~MK (T)   &  $0.521 \pm 0.003$ & 2 & 1.2 & $22.4\pm 0.5$ & $0.10 \pm 0.03 $ & 1.60 (?) & Period unknown\\
         2015~RG$_{36}$ (B) &  $0.31 \pm 0.01$   & 1 &  0.05       & $20.7 \pm 0.4$ & $0.21 \pm 0.05$ & 1.04 & Period unknown\\
(488693) 2003~WW$_{87}$ (B) &  $4.676 \pm 0.005$ & 3 & 0.11        & $17.0 \pm 0.3 $ & $1.1 \pm 0.3$ & 1.07 &Period unknown\\
(154589) 2003~MX$_{2}$ (B)  &  $2.3 \pm 0.2 $    & 2 & 0.04        & $16.7\pm 0.3$ & $1.3\pm 0.3$ & 1.03 & Uncertain period of about 1.6 h\\
         2024~WB (B)   &  $0.302 \pm 0.006 $& 1 &  0.03       & $21.1 \pm 0.3$ & $0.30 \pm 0.1$ & 1.03 & Period unknown\\   
(618350) 2021~PS$_{2}$ (B)  &  $0.27 \pm 0.02$   & 1 &  $0.06$     & $19.8 \pm 0.4$ & $0.5 \pm 0.2$ & 1.04 & Period unknown\\  
(9058) 1992~JB (B)     &  $7$ (?)           & 1 &  $> 0.34$   & $18.3 \pm 0.3$ & $0.6 \pm 0.1$ & - & Period unknown\\
(812769) 2005~VO$_{5}$ (B)  &  $1.065 \pm 0.001$ & 2 &   0.23      & --- & ---  & 1.20 & Period unknown.\\
(333284) 1999~PJ$_{1}$ (B)  &  ---               & - &   ---       & $18.4 \pm 0.2$ & $0.5 \pm 0.1$ & - & Period known, 6.2013 h\\
(152664) 1998~FW$_{4}$ (B)  &  ---               & - &   ---       & $19.7 \pm 0.3$ & $0.5 \pm 0.2$ & - & Period known, 17.38 h\\
        2025~FA$_{22}$ (B)  &  $13.1 \pm 0.2$    & 2 &   0.60      & $21.4 \pm 0.3$ & $0.2 \pm 0.1$ & 1.50 & Period unknown\\
(138205) 2000~EZ$_{148}$ (B)  &  $4.981 \pm 0.002$ & 3 &   0.17    & $15.5 \pm 0.2$ & $2.4 \pm 0.5 $ & 1.14 & Period unknown\\   
(843079) 2016~CB$_{32}$ (B)  &  (?)  & - &  $>0.04$   & $18.9 \pm 0.3$ & $0.7 \pm 0.3 $ & - & Period unknown\\
(52340) 1992~SY (B)     &  $3.1485 \pm 0.0005$ & 3 &  $0.10$  & $18.3 \pm 0.2$ & $0.5 \pm 0.1$ & 1.05 (?) & Period unknown\\
		\hline
	\end{tabular}
\end{table*}

\begin{figure}
    \centering
    \includegraphics[width=1.0\textwidth]{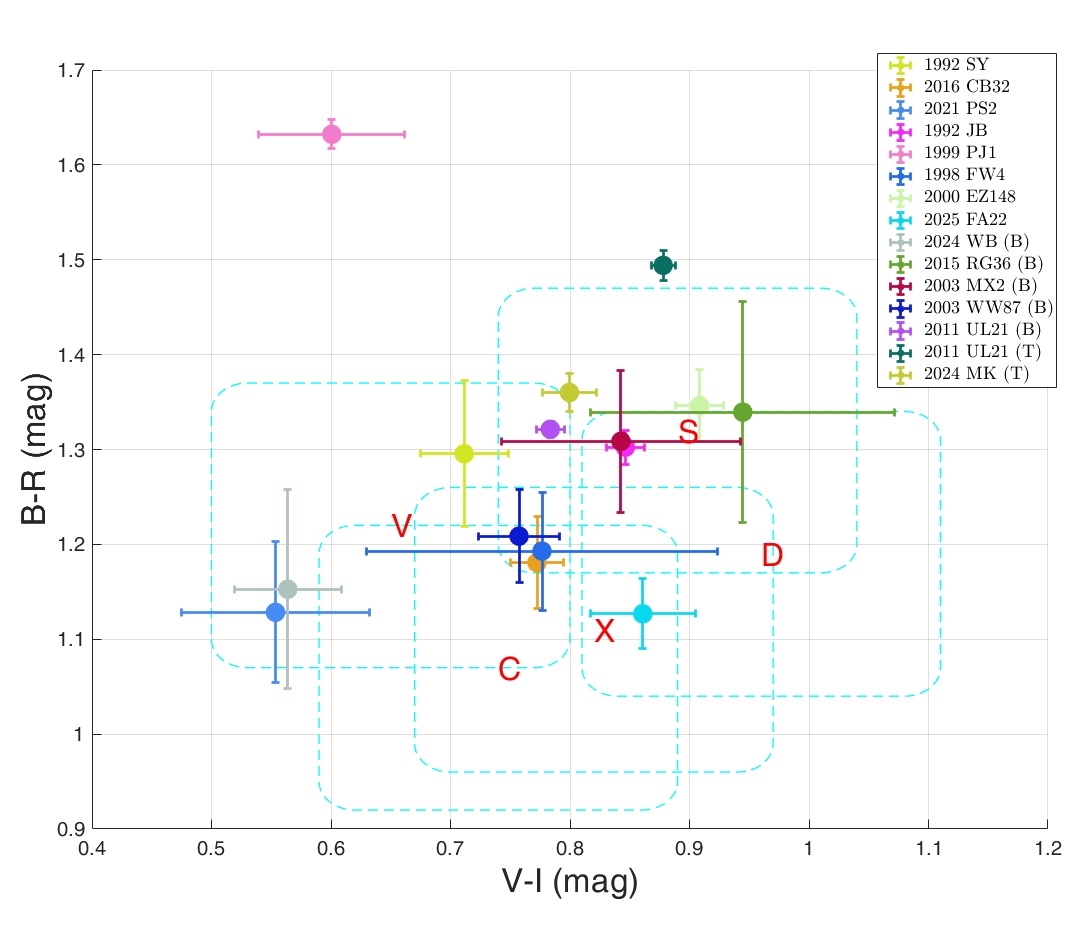}
    \caption{The position of the observed NEAs in a $B-R_c$ vs $V-I_c$ diagram with the positions of the main taxonomic classes indicated. The class boundaries are within 0.15 mag from the mean values given in \cite{Dandy2003}. The letter near the asteroid's name is B for BFOSC and T for TANDEM.}
    \label{fig:colors}
\end{figure}

\section{Results and Discussion}
\label{sec:dis}
The results about the color indices, taxonomic classification, effective diameter, amplitude, and rotation periods are collected in Table~\ref{tab:Neas} and Table~\ref{tab:Neas_period}. In Figure~\ref {fig:colors}, the observed NEAs are plotted in a $B-R_c$ vs $V-I_c$ diagram, while more extensive comments about each observed NEA, including the observational circumstances, the lightcurve, the color indices values, and the dimension estimate, are provided in the Appendix~\ref{sec:appendix}.\\
We have observed 15 near-Earth asteroids with estimated dimensions ranging from 0.1 to 2.4 km, for which we derived color indices, tentative taxonomic classifications based on reflectance curves, rotation periods, and approximate physical parameters, including the absolute $V$ magnitude, the effective diameter, and the minimum axis ratio. Both the $HG$ and $HG_1G_2$ models were used to estimate the absolute magnitude and diameter, with no significant differences found, within uncertainties, in the final results. The values in Table~\ref{tab:Neas_period} are obtained with the $HG_1G_2$ model. While the sample is small and selected based on observational constraints (e.g., brightness and visibility), it nevertheless provides a useful set of case studies for exploring the rotational and compositional properties of NEAs.\\

\subsection{Taxonomic classification and sample properties}
Color indices were obtained for 14 objects and used to infer taxonomic classes by comparing them with reference reflectance curves, see Section~\ref{sub:tax}. The majority of the sample appears consistent with S-complex asteroids, with additional objects compatible with C-complex, V-type, X-complex, and T-type. This distribution is broadly in line with previous surveys of NEAs, although the small size and observational selection of our sample prevent any statistically robust conclusions about population fractions. There are four asteroids observed at high phase angle $\alpha \geq 40^\circ$, 2011 UL$_{21}$, 2024 MK, 2003 WW87 and 1992 SY. The first three are classified as S-complex asteroids, so, based on Table~\ref{tab:slope_color}, they are little affected by phase reddening. The last asteroid is a V-type and may exhibit phase reddening, so the classification is uncertain. \\
Anyway, approximately 50\% of the asteroids belong to the S-complex, 14\% to the C-complex, 21\% to the X-complex, and 14\% to the V-type. The classification in Table~\ref{tab:Neas} is based on the reflectance curve and reports, in brackets, the probability estimated using the Monte Carlo technique. \\
As a comparison, we cite the recent results of \cite{Hromakina2021}, which found that among 55 observed NEAs, 43\% were S-complex, 19\% were X-complex, 16\% were C-complex, 12\% were D-types, and 6\% and 4\% were A- and V-types, respectively. Considering that our sample is much smaller, these differences appear reasonable. Expected taxonomy counts and their standard deviation of a subsample of a larger sample can be computed assuming sampling without replacement from a finite population, using the hypergeometric distribution \citep{Feller1968}. Taking as a reference the bigger NEAs sample, the expected S-complex asteroids in our sample is $S_{mean} = p\cdot n= 0.43\cdot 14\approx 6$ with standard deviation:

\begin{equation}
S_{SD}=\sqrt{np\left(1-p\right)\frac{N-n}{N-1}}\approx \sqrt{14\cdot0.43\left(1-0.43\right)\frac{55-14}{55-1}}\approx 2
\label{eq:SD}
\end{equation}

\noindent So, for S-complex asteroids in our sample, we can expect 4-8 asteroids of this class, equivalent to 28\%-60\% of the total. For C-complex asteroids, the expected mean value is 2.2, with a standard deviation given by Eq.~\ref{eq:SD} of 1.4, corresponding to a range of 7\% to 26\%, also in good agreement. Lastly, for X-complex, the expected percentage in our sample is between 10\% and 28\%, and so appears also in good agreement.\\
It is important to emphasize that taxonomic classifications based on broadband $BVR_cI_c$ photometry should be regarded as indicative rather than definitive. Therefore, the taxonomic assignments presented here should be interpreted as the best solutions within the adopted framework, pending confirmation by spectroscopic observations.

\begin{figure}
    \centering
    \includegraphics[width=1.0\textwidth]{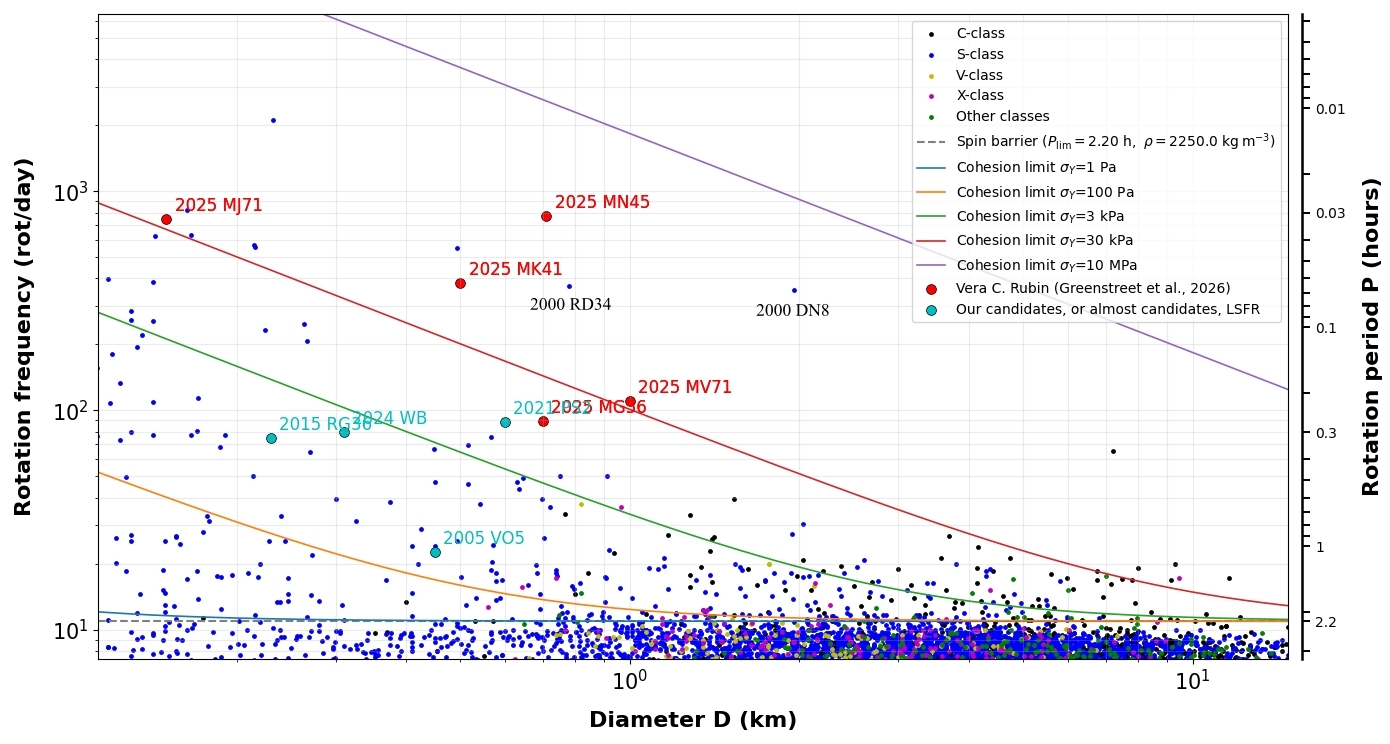}
    \caption{Plot of the frequency vs. diameter for all asteroids with known or estimated rotation period with quality code $U\geq 1$. File data from the ALCDB of 1 Oct 2023, the last release (\url{https://alcdef.org/}). The two near-Earth asteroids, 2000~RD$_{34}$ and 2000~DN$_{8}$ (both with $U=1$), are shown together with the five LSFR from Rubin observations (red) and the candidate LSFR or almost candidate (due to borderline diameter as for 2015~RG$_{36}$ and 2024~WB), which we have characterized at the Loiano Astronomical Station, managed by INAF-OAS in Italy (cyan). The curves for the rubble pile model with cohesion strength are from Eq.~\ref{eq:cohesive_strength} \citep{Sanchez2014}. Our candidates LSFR has a cohesive strength between $10^{2}$--$10^{4}$~Pa.}
    \label{fig:freq_diam}
\end{figure}

\subsection{Candidate super-fast rotators and cohesive strength}
\begin{figure}
    \centering
    \includegraphics[width=1.0\textwidth]{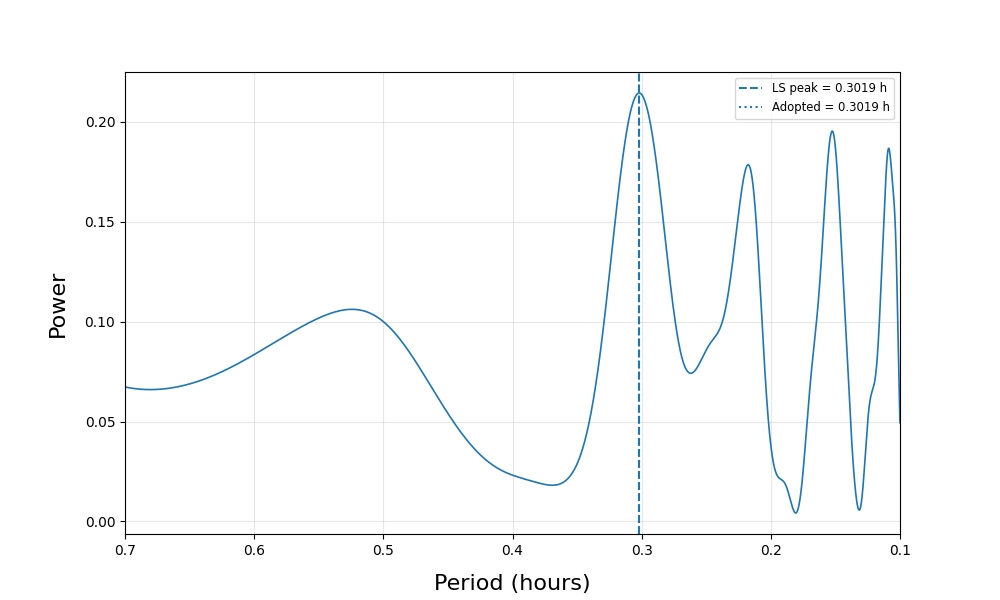}
    \caption{The Lomb-Scargle periodogram of 2024~WB shows the presence of two bimodal periods $P$ at 0.30 and 0.22 h, with the corresponding $P/2$ at 0.15 and 0.11 h, indices of a possible double period in the lightcurve. }
    \label{fig:2024_WB_LS}
\end{figure}

Rotation periods were determined or constrained for most targets using time-resolved photometry and period-search techniques with FALC and the Lomb-Scargle algorithms, as explained in Section~\ref{sub:period}. For a subset of objects, the observational time span is limited to a few hours or even a single session. In these cases, the derived periods should be considered provisional, as aliasing effects and ambiguities between solutions cannot be fully excluded, even if the false alarm probabilities are low.\\
Between the lightcurves of the observed NEAs, that of 2024~WB, and maybe also in 2021~PS$_{2}$, suggest the presence of multiple periodicities; see Fig.~\ref{fig:2024_WB_LS}. These features may indicate a binary system with asynchronous components with a proper rotation period or non-principal-axis rotation (tumbling), see Section~\ref{sec:2024WB} for more details. The further analysis presented in the appendix is based on phenomenological fits using sums of Fourier series adequate to explain the first hypothesized scenario, but does not constitute a physical model of tumbling rotation. Therefore, while this last interpretation is plausible, they remain speculative given the available data. A rigorous confirmation of tumbling states would require more extensive datasets.

For 5 asteroids, namely 2005~VO$_{5}$, 2015~RG$_{36}$, 2021~PS$_{2}$ (with some reservations, see Subsection~\ref{sec:2021PS2}), 2024~MK, and 2024~WB, have candidate rotation periods of less than about 1 h, although in some cases additional observations are required to confirm these values, while 2003~MX$_{2}$ has a period slightly superior to 2 h, see Fig.~\ref{fig:freq_diam}. The rotational properties of asteroids provide fundamental constraints on their internal structure, mechanical strength, and evolutionary history. \\
For kilometer-scale asteroids, rotation periods are longer than about 2.2 hours, corresponding to a frequency of about 11 rotations per day, while asteroids with dimensions smaller than 0.15 km can have higher frequencies. This limit defines the so-called cohesionless spin barrier \citep{Pravec2000}, which is the consequence of a cohesionless rubble-pile structure, in which asteroids are made up of collisional breakup fragments bound together by mutual gravitational force only. However, there are objects with diameters well greater than 0.15 km and rotation frequencies exceeding this limit that are classified as super-fast rotators (SFRs). These asteroids are of particular interest because they likely require non-negligible internal cohesion or a largely monolithic structure to remain intact \citep{Holsapple2007}. While a population of small SFRs is now well established, confirmed large SFRs (LSFRs) with diameters of 300–500 m or more remain rare and observationally challenging. Between the known NEAs, (29075) 2000~RD$_{34}$ and (331471) 2000~DN$_{8}$ have been reported as very large candidate LSFRs, see Fig.~\ref{fig:freq_diam}. Based on the absolute magnitude, the first has a diameter of about 1 km, while the second has a diameter of about 1.7 km; however, their rotation periods and lightcurve amplitudes are either poorly constrained or based on limited temporal coverage. Unfortunately, they are too faint targets to be in the range of our instruments.\\
Our asteroids with periods less than one hour can be considered candidate fast or super-fast rotators. However, given the uncertainties discussed above, especially for periods derived from limited datasets, these classifications should be regarded as tentative until confirmed by independent observations. The presence of candidate fast rotators with estimated diameters exceeding $\approx 0.15$ km may suggest that internal cohesion plays a role in maintaining structural integrity. In our case, the NEA 2024 MK is about 0.1 km in size and can be considered monolithic, whereas 2005~VO$_{5}$, 2015~RG$_{36}$, 2021~PS$_{2}$, and 2024~WB are larger than about 0.2 km, so a more complex rubble-pile model with cohesion forces, such as \cite{Sanchez2014}, is required to explain their existence. As an example, we applied it to 2015~RG$_{36}$ and 2021~PS$_{2}$, but consider that the amplitude of their light curves is low and the cycles covered are few, so aliasing and spurious periodogram peaks cannot be excluded. For this reason, we will also apply the cohesive rubble pile model to half- and double-rotation periods. \\
For 2015~RG$_{36}$, we estimated a diameter of $D=0.21 \pm 0.05$ km, and the spin barrier rate $\omega=2\pi/P$ for an asteroid with cohesive strength is given by \citep{Sanchez2014}: 

\begin{equation}
    \omega^2\leq \omega_0^2+\frac{\sigma_Y}{\rho r^2}
    \label{eq:cohesive_strength}
\end{equation}

\noindent In Eq.~\ref{eq:cohesive_strength}, $\rho\approx 2720 ~\text{kg}~\text{m}^{-3}$ is the S-complex mean density \citep{Carbognani2017}, $\omega_0 \approx 1.67\cdot 10^{-5}\cdot\sqrt{\rho}\approx 8.7\cdot 10^{-4} ~\text{s}^{-1}$ is the cohesionless spin-barrier rotation frequency, $\sigma_Y$ is the cohesive strength (force for the unit area),  $\omega \approx 5.6\cdot 10^{-3} ~\text{s}^{-1}$ is the observed rotation frequency and $r\approx 115 ~\text{m}$ is the asteroid effective radius. With these data from Eq.~\ref{eq:cohesive_strength}, we obtain an order-of-magnitude estimate of $\sigma_Y\geq \rho r^2\left(\omega^2-\omega_0^2\right)\approx 1.1\cdot 10^3$ Pa, subject to uncertainties in size, density, and shape assumptions. Doubling the rotation period, the cohesive strength drops to 260 Pa, while at half the period, the strength rises to $4.5\cdot 10^{3}$ Pa. The same model can be applied to 2021~PS$_{2}$ ($D=0.5 \pm 0.2$ km). This NEA belongs to the C-complex, so $\rho\approx 1330 ~\text{kg}~\text{m}^{-3}$ and $\omega_0\approx 6.1\cdot 10^{-4} ~\text{s}^{-1}$. With $r\approx 300 ~\text{m}$ and $\omega \approx 6.5\cdot 10^{-3} ~\text{s}^{-1}$, we get $\sigma_Y\geq 5\cdot 10^3$ Pa. Also in this case, doubling the period, the cohesion drops to $1.2\cdot 10^3$ Pa, while with a half period, it rises to $2\cdot 10^4$ Pa. \\
These values should be regarded as indicative rather than definitive, given the model's simplifying assumptions. Anyway, all these values appear physically acceptable considering that the strong cohesion of lunar rocks is about $\sigma_Y\approx 3\cdot 10^3$ Pa \citep{Sanchez2014}, while the global strength of the little Chelyabinsk asteroid, about 20 m in diameter, was estimated from the main fragmentation height to be about $2\cdot 10^6$ Pa \citep{Carbognani2024}. This difference in strength between the Chelyabinsk asteroid and relatively large asteroids such as 2015~RG$_{36}$ and 2021~PS$_{2}$ is probably due to the former being monolithic, while the latter were not.\\
The relevance of these observations is further emphasized by the recent discovery of large fast-rotating asteroids in the main belt by the Vera C. Rubin Observatory \citep{Greenstreet2026}. From Rubin data, there are 16 super-fast rotators with respect to 76 asteroids with well-established periods, with rotation periods ranging from approximately 13 minutes to 2.2 hours, and 3 ultra-fast rotators that complete a full spin in under 5 minutes for a total of 19 SFRs, about 26\% of the total. These objects demonstrate that rapid rotation is not limited to small bodies and may occur across a wider size range than previously recognized.\\
If an LSFR asteroid with a non-zero probability of hitting Earth were discovered, an estimate of its minimum strength could be obtained. These constraints on internal strength and rotational stability are potentially relevant to impact-risk assessment, as they may affect the structural response of NEAs to tidal forces, atmospheric entry, and mitigation attempts.

\section{Conclusions}
\label{sec:end}
We presented the results of a dedicated photometric study of near-Earth asteroids (NEAs) carried out between 2024 and 2026, based on broadband $BV R_{c} I_{c}$ photometry and time-resolved lightcurve observations. The primary goals of this work were to determine color indices, infer taxonomic classifications, and constrain the rotational properties and physical characteristics of a sample of NEAs that had never been studied before.\\
Reliable color measurements were obtained for all targets, with typical uncertainties that allow a first-order taxonomic assessment using the reflectance-curve fitting. The sample is dominated by S-complex asteroids, with additional representatives of the C-complex, X-complex, and V-type. This distribution is broadly consistent with previous photometric and spectroscopic surveys of the NEA population. \\
Rotation periods were determined or constrained for most objects in the sample. Some asteroids were found to rotate faster than the classical cohesionless spin barrier at $  P\simeq 2.2$~h. In particular, multiple objects with effective diameters of several hundred meters exhibit rotation periods shorter than 1 hour, potentially placing them in the large super-fast-rotator regime, pending confirmation of their rotational properties. These results are consistent with the growing evidence that rapid rotation may not be limited to very small NEAs.\\
By comparing observed rotation frequencies and estimated sizes with theoretical models of rubble-pile asteroids that include internal cohesion, we find that the rapid rotation of the larger objects is difficult to reconcile with a purely cohesionless model, within the uncertainties of the derived parameters. Modest cohesive strengths, on the order of $10^{2}$--$10^{4}$~Pa, are sufficient to explain the observed properties, consistent with values inferred for granular materials and regolith-dominated bodies. This supports the interpretation that many fast-rotating NEAs are gravitational aggregates with non-negligible internal strength rather than monolithic bodies.\\
Overall, this work highlights the continued relevance of targeted photometric observations for characterizing the physical and rotational properties of NEAs. Such studies remain essential for improving our understanding of asteroid internal structure and evolution, with direct implications for planetary defense, in the event that an asteroid of non-negligible size and with a certain probability of impacting the Earth were to be characterized. We have also decided to extend the survey, so a second set of NEAs will be observed in the coming months.

\section*{Acknowledgements}
This paper uses data collected at the 1.52-m ``G. D. Cassini'' Telescope, managed by INAF-Osservatorio Astrofisico e Scienza dello Spazio di Bologna. Thanks to the ``Cassini'' technical staff for their support during observations. The authors would like to thank the staff of the San Marcello Pistoiese Astronomical Observatory (IAU 104) for their assistance with follow-up observations.

\section*{Data Availability}
The data underlying this paper will be shared on reasonable request to the corresponding author. The lightcurves was uploaded to ALCDEF database.

\bibliographystyle{cas-model2-names}
\bibliography{neas_biblio}{}

\appendix

\section{Results on individual asteroids}
\label{sec:appendix}
In this appendix, the observational circumstances and the main results obtained for each observed NEA are reported in chronological order. 

\subsection{(415029) 2011~UL$_{21}$}
\label{sec:2011UL21}
Asteroid 2011~UL$_{21}$ is in an Apollo-type orbit and a Potentially Hazardous
Asteroid (PHA). This object was observed during the close approach on June 27, 2024. This NEA was first observed for about one hour on Jun 30, 2024, with TANDEM using the $BVR_cI_c$ filters when it was already receding from Earth but still bright at mag +12.1 and with a moderate angular velocity of 23 arcsec/min. It was impossible to follow the approach phase because it was not visible from IAU 598. During this session, we found the following color indices: $B-V=0.93 \pm 0.01$, $V-R_c=0.56 \pm 0.01$, and $V-I_c=0.88\pm 0.01$ mag. A second photometric run of about a quarter of an hour was done on July 12 with the BFOSC, and we found: $B-V=0.85 \pm 0.01$, $V-R_c=0.47 \pm 0.01$, and $V-I_c=0.78\pm 0.01$ mag. The two sets of color indices (see also Table~\ref{tab:Neas}) are not identical, and this could indicate a macroscopic variation of colors on the asteroid's surface, but both are compatible with an S-complex asteroid (see Fig.~\ref{fig:colors}). The same classification as S-complex asteroid is obtained using the reflectance curve given by Eq.~\ref{eq:reflect}.\\
Previous measurements of the 2011 UL$_{21}$ colour indices were made by \cite{Hromakina2021} with $B-V=0.81 \pm 0.03$, $V-R_c=0.53 \pm 0.02$ and $V-I_c=0.96\pm 0.04$ mag and \cite{Novakovic2024} with $B-V=0.89 \pm 0.06$, $V-R_c=0.48 \pm 0.07$ mag. As we can see, both $B-V$ and $V-R_c$ agree with BFOSC's value, but they differ slightly from TANDEM's. On the other hand, the index $V-I_c$ of \cite{Hromakina2021} is nearer to TANDEM than the BFOSC value. This increases the likelihood of macroscopic color variations on the asteroid's surface.\\
During observation periods, the lightcurves of this NEA remained substantially flat, and this is due to a low amplitude lightcurve because Goldstone imaging on Jun 27, 2024, revealed that 2011 UL$_{21}$ is a binary system and places an upper bound of 2.5 h on the rotation period of the primary, with an equatorial diameter of about 1500 m \footnote{\url{https://echo.jpl.nasa.gov/asteroids/june2024.goldstone.planning.html}}.\\
With the BFOSC photometric observations, the mean absolute magnitude from Eq.~\ref{eq:HG1G2} yields $H_V=16.3 \pm 0.3$ mag, whereas with TANDEM, we obtain $H_V=16.4 \pm 0.3$ mag. As we can see, the two magnitude absolute values are consistent and in agreement with the value given by the Minor Planet Center ($H=15.93$), which, however, incorporates astrometric observations from different surveys with varying photometric bands and calibration levels, as noted in \cite{Hoffmann2024}. Considering that an S-complex asteroid has a geometric albedo $p_V=0.208 \pm 0.079$ \citep{Usui2013}, from Eq.~\ref{eq:diameter} with BFOSC's data, we get a mean diameter of about $D=1.6 \pm 0.4$ km, while with TANDEM's data, $D=1.5 \pm 0.3$ km. These values are comparable with the NEOWISE value \citep{Masiero2020} that, with a thermal model fit, gives $H=15.8$ mag for the absolute magnitude (with the fixed value $G=0.15$) and $D=2.3\pm 0.7$ km for the diameter, and are also in excellent agreement with the Goldstone radar result.

\begin{figure}
    \centering
    \includegraphics[width=1.0\textwidth]{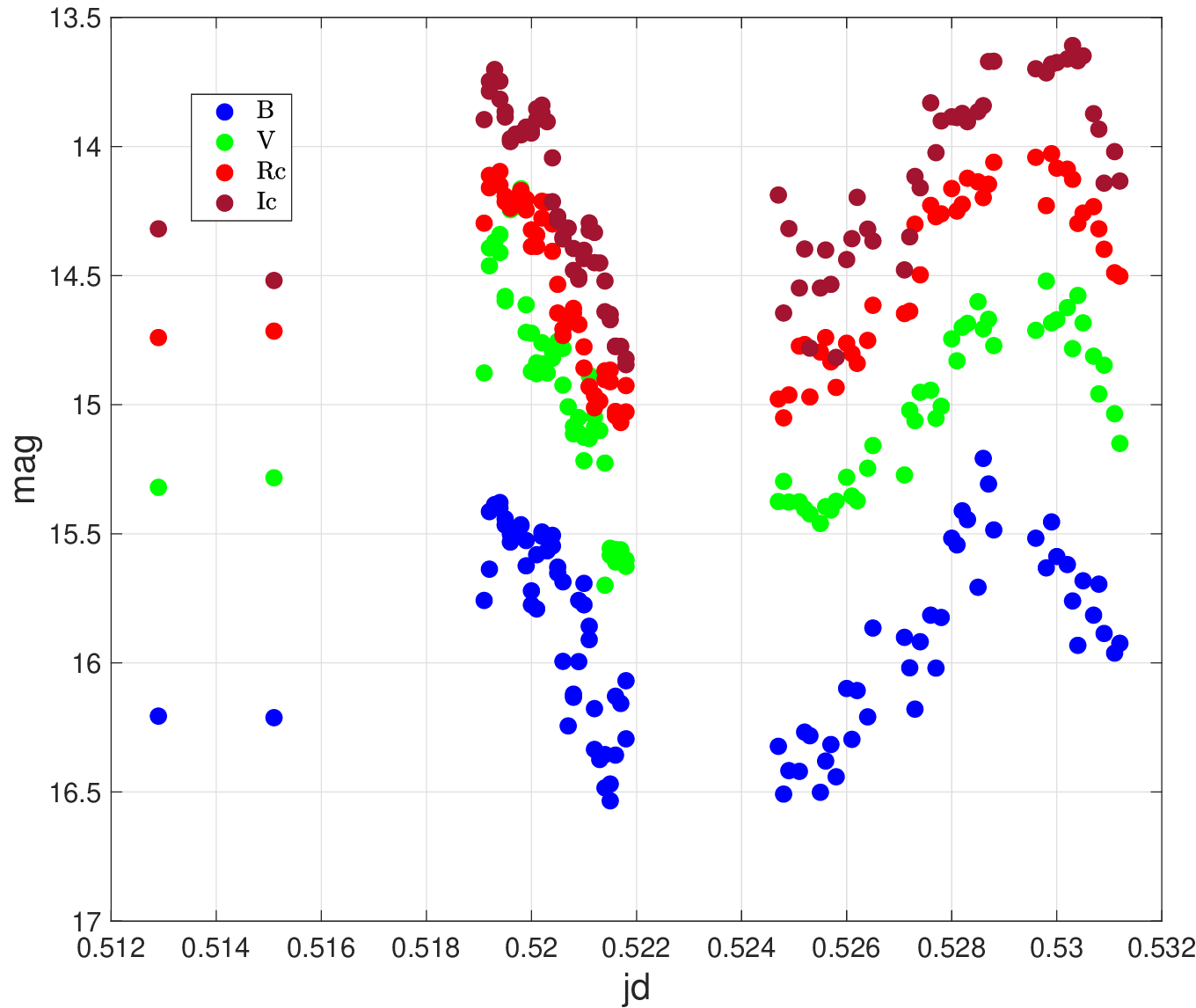}
    \caption{The four $BVR_{c}I_{c}$ lightcurves for 2024~MK taken with TANDEM during the night of the 30 Jun 2024. On the abscissa is the fraction of the Julian date of the day 2460492.}
    \label{fig:2024MK_colours}
\end{figure}

\begin{figure}
    \centering
    \includegraphics[width=1.0\textwidth]{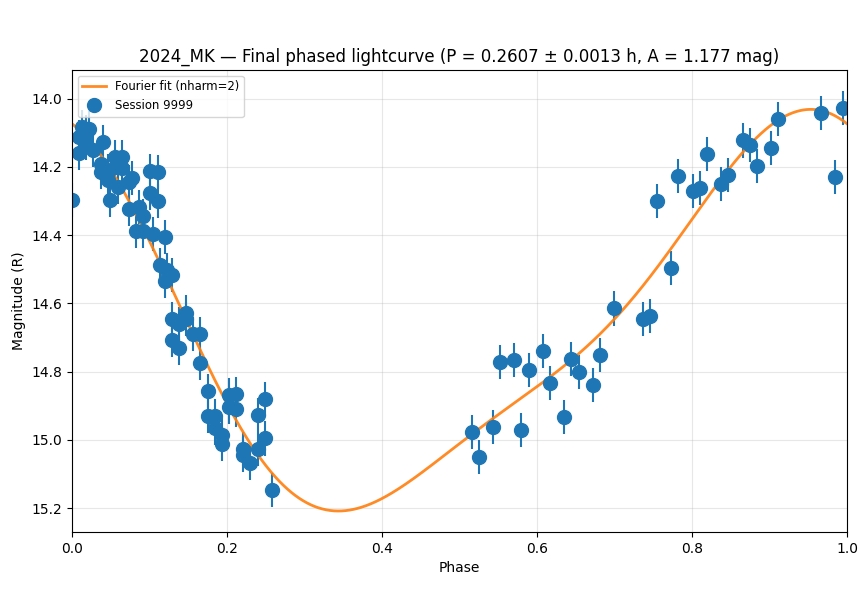}
    \caption{The phased lightcurve with the Lomb-Scargle algorithm and n = 2 for 2024~MK shows a period of $0.2607 \pm 0.0013$ h, corresponding to a monomodal phased lightcurve. Assuming a bimodal lightcurve as the most common feature for asteroids, the true period is double, so $0.521 \pm 0.003$ h. The lightcurve amplitude is 1.18 mag.}
    \label{fig:2024MK_pl}
\end{figure}

\subsection{2024~MK}
\label{sec:2024MK}
Asteroid 2024~MK was observed during the close approach of 29 Jun 2024. Also, for 2024~MK, the approach phase was not observable, so we used TANDEM to do $BVR_cI_c$ photometry on the night of 30 Jun 2024 with an apparent magnitude of about +14.5 mag. The observations covered a half-hour period during which the TANDEM system captured 94 images in 4 colors, though not uniformly spaced in time. The lightcurves $BVR_cI_c$ as a function of the fraction of the Julian date are reported in Fig.~\ref{fig:2024MK_colours}. As shown, photometric variability due to the asteroid's rotation is evident. With the Lomb-Scargle algorithm, the monomodal lightcurve (with FAP=0) has a rotation period $P=0.261 \pm 0.0013$ h (see Fig.~\ref{fig:2024MK_pl}), so the full period of the lightcurve with two maxima and two minima - as expected from an elongated asteroid - is $P\approx 0.521 \pm 0.003$ h with an amplitude of about 1.18 mag. The same result is given by the FALC algorithm. Apart from the measurement uncertainties, there are no areas of the asteroid's surface with markedly different colors, and the value reported for the color indices in Table~\ref{tab:Neas} is the average of all the observed color indices values with the associated uncertainty of the mean: $B-V=0.91 \pm 0.02$, $V-R_c=0.45 \pm 0.02$, and $V-I_c=0.80 \pm 0.02$ mag.\\
From mean magnitude measurements in $V$, as already done for 2011 UL$_{21}$, we obtain from Eq.~\ref{eq:HG1G2} an average absolute magnitude $H_V=22.4\pm 0.5$ mag, consistent with that given by the Minor Planet Center ($H=22.01$). Since from reflectance 2024~MK is an S-complex asteroid, the effective diameter from Eq.~\ref{eq:diameter} is $D=0.10 \pm 0.03 $ km. The same classification results from color indices. The asteroid 2024~MK was observed by Goldstone radar on June 29, 30, and July 1, 2024\footnote{\url{https://echo.jpl.nasa.gov/asteroids/june2024.goldstone.planning.html}}. The first radar results indicate a long axis of approximately 130 meters, characterized by hills, ridges, concavities, and boulders, with a tumbling rotation state exhibiting periods of 0.37 and 0.50 hours. These values align with our findings, except for the tumbling state, which, due to the brevity of the observation, was not detected.

\subsection{2015~RG$_{36}$ }
\label{sec:2015RG36}
\begin{figure}
    \centering
    \includegraphics[width=1.0\textwidth]{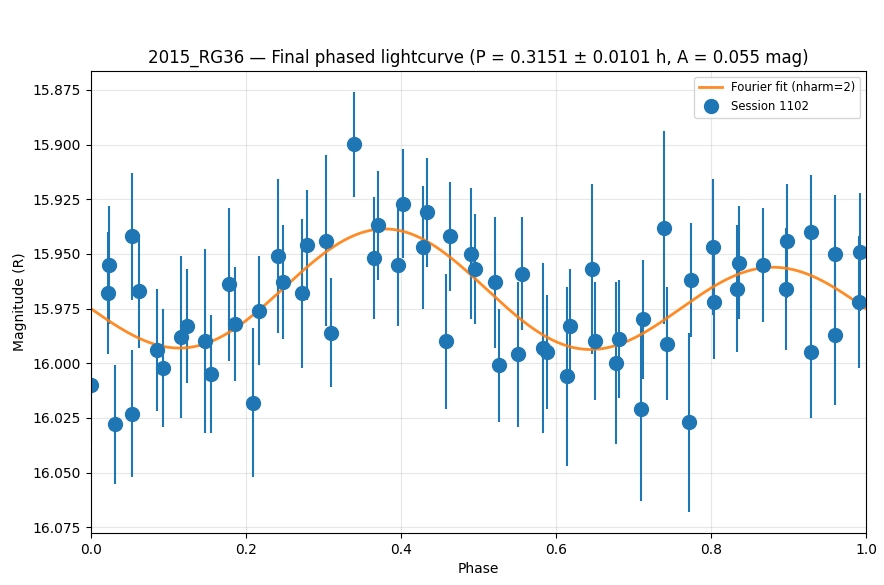}
    \caption{The bimodal phased lightcurve of NEA 2015~RG$_{36}$ with the Lomb-Scargle algorithm. The resulting rotation period is about 19 minutes, well below the classical spin barrier value.}
    \label{fig:2015RG36}
\end{figure}

The asteroid 2015~RG$_{36}$ is in an Amor-type orbit. This NEA was observed on 11 Nov 2024 with BFOSC during the separation phase from Earth after the minimum distance was reached on 9 Nov 2024. The telescope session took place with the Moon in phase 0.77, waxing, and with an initially clear sky that worsened during the evening, gradually clouding over. Despite this, a dense photometric session between 20:30 and 21:30 UTC in the $R_c$ band with 30 s exposures highlighted a bimodal lightcurve with a Lomb-Scargle best rotation period of $P=0.31 \pm 0.01$ h and an amplitude $A=0.05$ mag, see Fig.~\ref{fig:2015RG36}. In the LS periodogram, a second peak is also present for the semiperiod, as expected for a real signal. Using the Lomb-Scargle algorithm, the FAP with the bootstrap method is 1\%. A further photometric session in $R_c$ was conducted the following evening; however, the weather was not optimal, and the resulting lightcurve cannot be used to improve the period. The color indices obtained at the beginning of the session with a clear sky, by comparing the asteroid's magnitude with the Landolt PG0220 field, are as follows: $B-V=0.81 \pm 0.06$, $V-R_c=0.53 \pm 0.1$, and $V-I_c=0.94 \pm 0.1$ mag. The reflectance values are consistent with those of an S-complex asteroid. Also, as cited in \cite{Perna2018}, spectroscopic observations classify this asteroid in the same manner. In the case of 2015~RG$_{36}$, the estimated diameter based on S-complex mean albedo is $D=0.21 \pm 0.05$ km.

\subsection{(488693) 2003~WW$_{87}$}
\label{sec:2003WW87}
\begin{figure}
    \centering
    \includegraphics[width=1.0\textwidth]{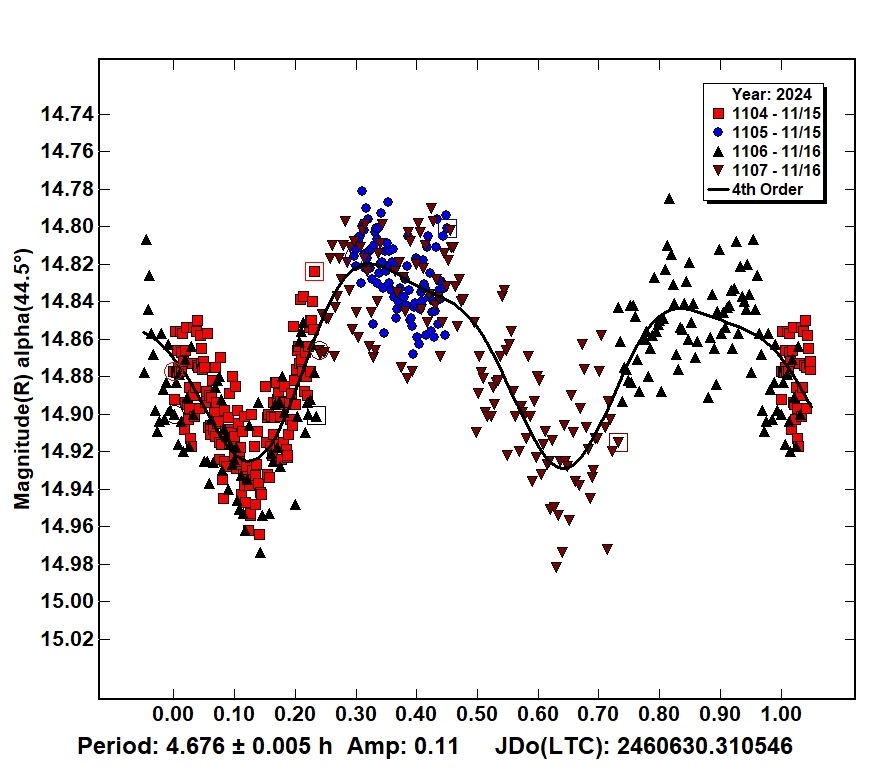}
    \caption{The phased lightcurve of NEA (488693) 2003 WW$_{87}$. The sessions 1104 and 1105 are from the ``G.D. Cassini'' telescope, while sessions 1106 and 1107 are from IAU 104. For ease of reading, the error bars of the photometric points have been removed: for IAU 598, they are $\pm 0.02$ mag, while for IAU 104, they are $\pm 0.05$ mag. }
    \label{fig:2003WW87}
\end{figure}

The Amor-type (488693) 2003~WW$_{87}$ was observed on 15 Nov 2024 with BFOSC during a full Moon night, but with a clear sky. The asteroid was about $55^\circ$ from the Moon, but it is bright (Table~\ref{tab:Neas}), so there is not a great disturbance from the light of our satellite. The color indices were determined by comparing the $BVR_cI_c$ images with the Landolt field PG2213 and are $B-V=0.77 \pm 0.05$, $V-R_c=0.43 \pm 0.02$, and $V-I_c=0.76\pm 0.03$ mag. The reflectance values appear most compatible with an S-complex asteroid.\\
The dense photometry in $R_c$ was divided into two sessions because, with an angular speed of about 10.7 arcsec/minute in a nearly south-to-north direction, the asteroid took about one hour to traverse the entire FoV of the BFOSC. With these first two sessions, the lightcurve showed a maximum and a minimum, but it was incomplete. For this reason, an auxiliary photometric session was conducted by IAU 104 on 16 Nov 2024 for almost 5 hours with a clear filter. Combining all the sessions, the most probable rotation period is $4.676\pm 0.005$ h with an amplitude of 0.11 mag, see Fig.~\ref{fig:2003WW87}. From the mean magnitude values in $V$ and using the mean value for the S-complex, we find $H_V=17.0 \pm 0.3 $ mag and $D=1.1 \pm 0.3$ km.

\subsection{(154589) 2003 MX$_{2}$}
\label{sec:2003MX2}
\begin{figure}
    \centering
    \includegraphics[width=1.0\textwidth]{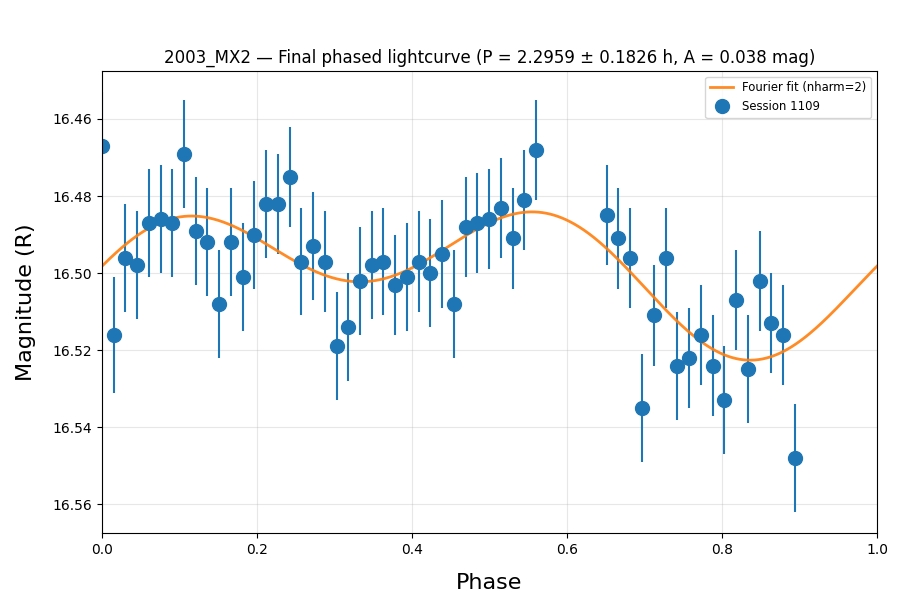}
    \caption{The phased lightcurve of NEA 2003~MX$_{2}$ with the best period found with the Lomb-Scargle algorithm ($n=2$). The bimodal rotation period is $2.3 \pm0.2$ h, with an amplitude of about 0.04 mag.}
    \label{fig:2003MX2}
\end{figure}

The NEA 2003~MX$_{2}$ is on an Amor orbit and was observed with BFOSC on December 2, 2024, during a night with a clear sky and no Moon. $BVR_cI_c$ observations were made to determine the color indices using Landolt PG2213 as a comparison field. The indices were as follows: $B-V=0.89 \pm 0.09$, $V-R_c=0.41 \pm 0.02$ and $V-I_c=0.84 \pm 0.1$ mag. The reflectance values are compatible with an S-complex asteroid.\\
A 2-hour photometric session with an $R_c$ filter was then started to determine the rotation period. In the JPL Small-Body Database, a period was indicated, even if uncertain, of about 1.611 h, a value lower than the cohesionless spin barrier that would not be expected for an asteroid of about 1 km in diameter. The data from our session show a low-amplitude lightcurve, about 0.04 mag, with a bimodal period of $2.3 \pm 0.2$ h, a value in line with the spin barrier (see Fig.~\ref{fig:2003MX2}). Based on the Lomb-Scargle algorithm, the false-alarm probability is about zero. If, as in the previous cases, the absolute magnitude $H_V$ is estimated from the $HG_1G_2$ phase function and the measured $V$ magnitudes, we find $H_V=16.7\pm 0.3$ mag. With this value, using the class-mean geometric albedo, we find $D=1.3\pm 0.3$ km, in good agreement with NEOWISE, which gives $D=1.1 \pm 0.3$ km.

\subsection{2024~WB}
\label{sec:2024WB}
\begin{figure}
    \centering
    \includegraphics[width=0.8\textwidth]{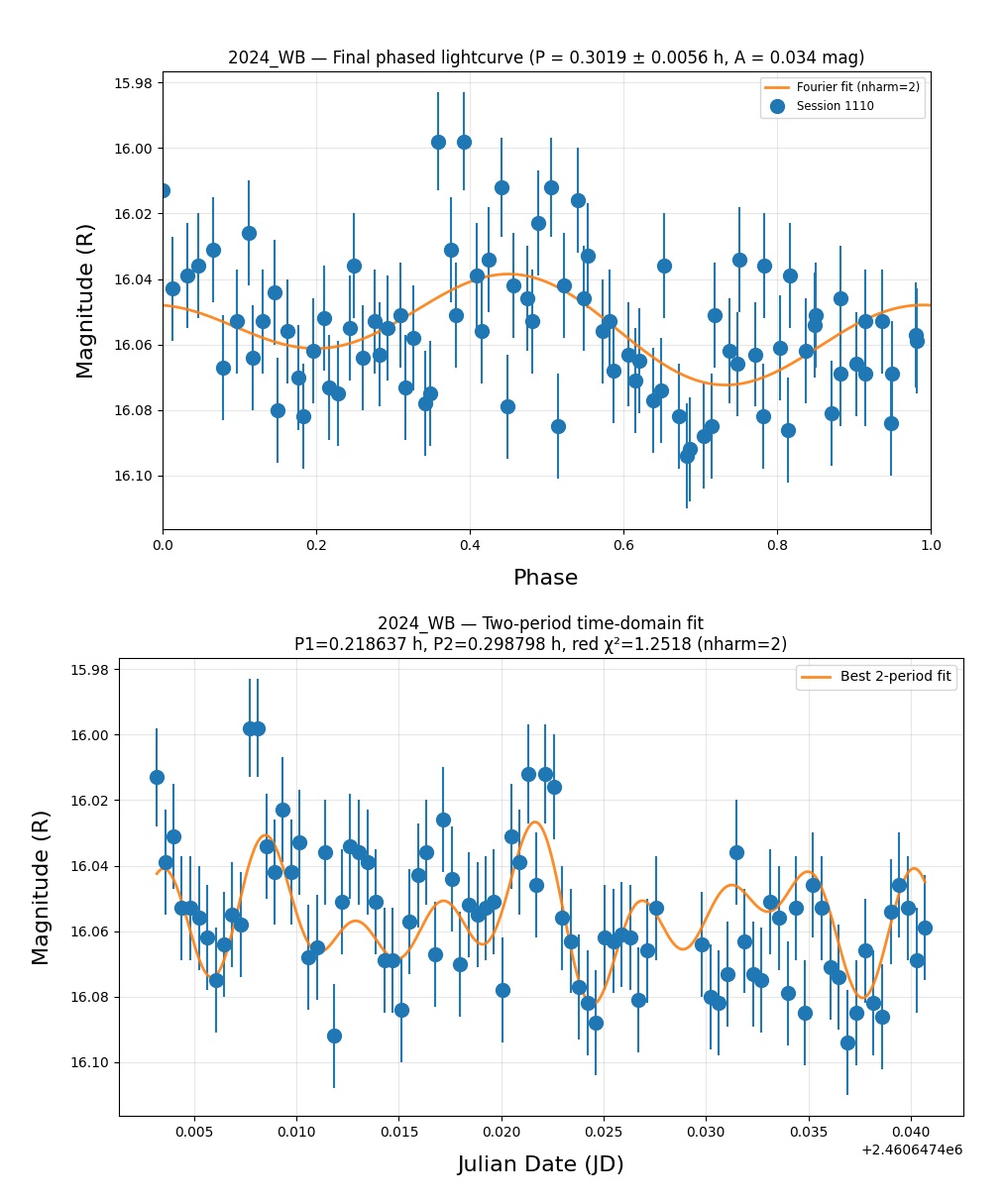}
    \caption{The lightcurve of the NEA 2024 WB. Top: best phased single period lightcurve with Lomb-Scargle analysis. Bottom: phenomenological least square best fit with a sum of two Fourier series in the range 0.1-0.5 h with periods of about 0.22 h and 0.3 h.}
    \label{fig:2024WB_Period}
\end{figure}

Asteroid 2024~WB is a PHA in an Apollo orbit and was observed with BFOSC also on December 2, 2024, after 2003~MX$_{2}$. In this case, the Landolt field PG0220 was used as a comparison for photometric calibration. The colour indices were as follows: $B-V=0.83 \pm 0.1$, $V-R_c=0.32 \pm 0.04$ and $V-I_c=0.56 \pm 0.04$ mag. Computing the reflectance and the probabilities of the taxonomic classes, the asteroid is consistent with a C-complex object, with a probability of 84\%.\\
The subsequent photometric session with the $R_c$ filter to determine the rotation period lasted one hour, and the results were interesting. With the FALC algorithm, there are four minima of almost comparable depth in the period spectrum, corresponding to periods of $0.311 \pm 0.005$ h with amplitude 0.04 mag and $0.231 \pm 0.004$ h with amplitude 0.03 mag, with their respective half-periods. The same scenario emerged from the Lomb-Scargle analysis: in the periodogram, there are two nearly equal peaks corresponding to a bimodal lightcurve with periods of $0.302 \pm 0.006$ h and $0.218 \pm 0.006$ h, with the first slightly higher than the second. The corresponding false-alarm probabilities are 1\% and 3.5\%, so the first period appears more reliable than the second. In the single-period phased lightcurve, some points appear as outliers from the best-fit curve, so we conducted an empirical search for a phenomenological double period in the 0.1-0.5 h interval by fitting a sum of two Fourier series, each with its own period. We found that the best couple that minimizes the $\chi^2$ is 0.22 h and 0.3 h, as shown in Fig.~\ref{fig:2024WB_Period}. These two periods may indicate that asteroid 2024~WB is an asynchronous binary system, with each component of comparable size having its own rotation period. Alternatively, the presence of these two periods could be due to the asteroid's tumbling state, although other interpretations (e.g., aliasing or noise-driven features) cannot be ruled out.\\
Most asteroids rotate at a constant rate around a direction fixed in space (pure spin state). This condition requires that the angular momentum vector \textbf{L} and the angular velocity vector $\omega$ are parallel along one of the body's three principal inertial axes (PIAs). However, for some asteroids in a most general rotation state, \textbf{L} and $\omega$ are not parallel with one another or with the body’s PIAs \citep{Paolicchi_2002}. This condition is known as non-principal-axis rotation (NPAR), and asteroids in this state are called tumbling asteroids. The tumbling state may result from a collision with an asteroid, a planet flyby, or radiation forces such as YORP. Interestingly, as in our case, tumbling asteroids show two frequencies in the lightcurve, but the decomposition in this case is more complex, see \cite{Pravec_2005}. Unfortunately, due to the low amplitude lightcurve, the data cannot distinguish between binarity, tumbling, aliasing, or noise.\\
As usual, the $H_V$ magnitude from calibrated $V$ value is $H_V=21.1 \pm 0.3$ mag, and the estimated diameter for C-complex is $D=0.3 \pm 0.1$ km. 

\subsection{(618350) 2021~PS$_{2}$}
\label{sec:2021PS2}
\begin{figure}
    \centering
    \includegraphics[width=0.8\textwidth]{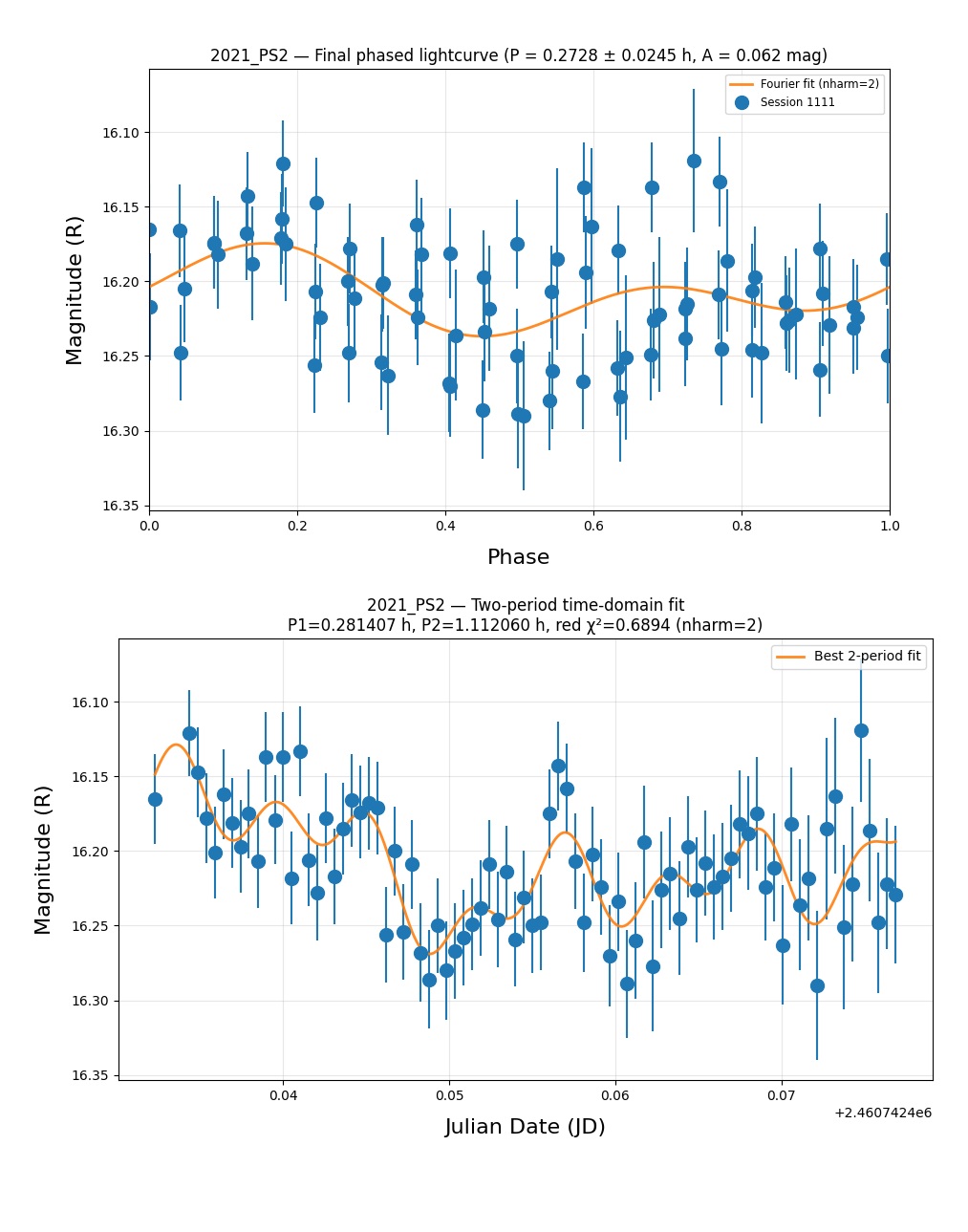}
    \caption{The lightcurve of NEA 2021~PS$_{2}$ obtained with a one-hour session. Up: phased lightcurve with a single best period of about 0.27 h. Bottom: lightcurve decomposition with the two least-square best periods of 0.28 and 1.11 h.}
    \label{fig:2021PS2}
\end{figure}

Asteroid (618350) 2021~PS$_{2}$ is an Apollo-PHA observed with BFOSC on March 7, 2025. In this case, the Moon was near the zenith, with an illuminated phase of about 0.63. The asteroid was at an angular distance of about $46^\circ$, so the moonlight disturbed it slightly, and the signal-to-noise ratio was 40 due to the diffuse light. The Landolt field PG0918 was used for photometric calibration. The colour indices of 2021~PS$_{2}$ result in the following: $B-V=0.82 \pm 0.1  $, $V-R_c= 0.27 \pm 0.01 $ and $V-I_c= 0.53 \pm 0.08 $ mag. Using the reflectance values from Eq.~\ref{eq:reflect}, the asteroid is consistent with the C-complex with a probability of about 88\%. The estimated absolute magnitude of 2021~PS2 is $H_V=19.8 \pm 0.4$ mag, and the effective diameter for the taxonomic class is $D=0.5 \pm 0.2$ km.\\
After the color indices, we conducted a dense photometric session using the $R_c$ filter for approximately 1 hour; a sudden cloud cover prevented us from continuing. The Lomb-Scargle analysis revealed a bimodal lightcurve with a periodicity of $p=0.27 \pm 0.02$ h and an amplitude of 0.06 mag (see Fig.~\ref{fig:2021PS2}). In the LS periodogram, there is also a secondary peak at $p/2$; however, the main peak power is low, and the FAP with the bootstrap method is near 10\%, so the suggested period does not seem very significant. For this reason, it appears with a question mark in Table~\ref {tab:Neas_period}. \\
The scatter of the points around the best-fit curve appears high, exceeding the photometric uncertainty, as for 2024~WB. For this reason, we conducted another empirical search for a phenomenological double period in the 0.1-2 h interval by fitting a sum of Fourier series, each with its own period. We found a local $\chi^2$ minimum with the pair of periods 0.28 and 1.11 h that significantly lowers the reduced $\chi^2$, from 1.25 for the single period to 0.69, see Fig.~\ref{fig:2021PS2} bottom. This can indicate that the asteroid is tumbling, although, given the limited data coverage, it is not possible to determine its individual values. An alternative scenario is a binary system with different periods for each component, but unfortunately, as for NEA 2024~WB, the data cannot distinguish among binarity, tumbling, aliasing, or noise. More photometric data will be necessary for this target.

\subsection{(9058) 1992~JB}
\label{sec:1992JB}
Asteroid (9058) 1992~JB is on an Apollo-type orbit and was observed with BFOSC on April 30 and May 26, 2025. The last session focused on refining the rotation period. The sky was clear in both sessions, with no moon visible. For the color $BVR_cI_c$ calibration on April 30, the Landolt field PG0918 was used. We also conducted a photometric session, 1.3 hours long with the $R_c$ filter, to determine the spin period, which was extended by another 2.8 hours in the second session.\\
Unfortunately, the two dense photometric sessions were not sufficient to unambiguously determine the rotation period. The semi-period could be 3.4 hours, in which case the complete rotation period, assuming a lightcurve with two maxima and two minima, would be about 7 hours.\\
The colour index are the following: $B-V=0.84 \pm 0.02$, $V-R_c=0.46 \pm 0.01$ and $V-I_c=0.85 \pm 0.02$ mag. The corresponding reflectance curve is compatible with the S-complex. So, the absolute $V$ magnitude is $H_V=18.3 \pm 0.3$ mag, and the corresponding diameter is $D=0.6 \pm 0.1$ km.

\subsection{(812769) 2005~VO$_{5}$}
\label{sec:2005VO5}
\begin{figure}
    \centering
    \includegraphics[width=0.8\textwidth]{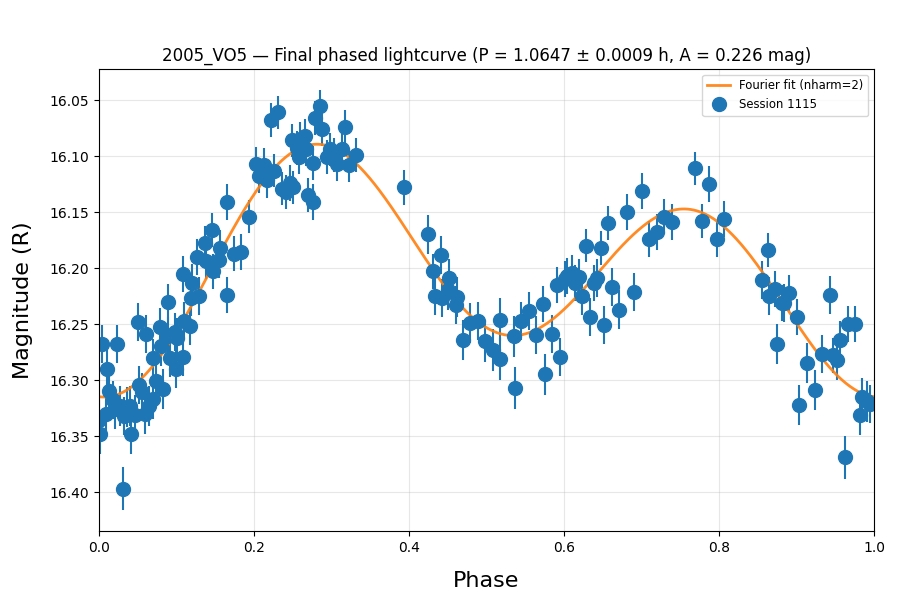}
    \caption{The phased lightcurve of NEA~2005 VO$_{5}$ obtained with a 2.5 hour session. The best bimodal rotation period from Lomb-Scargle with $n=2$ is about 1.065 h. However, with $n=4$, the best period is double, with a lightcurve with 4 maxima and 4 minima.}
    \label{fig:2005VO5}
\end{figure}

The PHA on Apollo orbit 2005~VO$_{5}$ was observed with BFOSC on the night of June 27, 2025, with stable, clear skies and no Moon. The asteroid's height above the horizon was just over $30^\circ$, the minimum value that the Cassini telescope can reach, and we chose to do a dense photometric session with an $R_c$ filter to determine its rotation period. Due to its rapid proper motion toward southern declinations (approximately 3.4 arcsec/minute), a session to determine its color indices was not conducted; unfortunately, this NEA will not pass Earth's proximity again until 2062. \\
The asteroid was located in the constellation Serpens, in a dense star field. To facilitate subsequent photometry, the BFOSC was used in bin 1 mode to better separate the background stars from the asteroid. A total of 240 images were taken with a 30-second exposure time using the $R_c$ filter for 2.5 hours. However, images in which the asteroid was superimposed on background stars were deleted to avoid altering the lightcurve. The phased lightcurve with Lomb-Scargle exhibits a bimodal structure, featuring two maxima and two minima, a period of $1.065 \pm 0.001$ hours, and an amplitude of about 0.23 mag (see Fig.~\ref{fig:2005VO5}). The peak power is high, the bootstrap false-alarm probability is near zero, and the LS periodogram also shows a $p/2$ peak. However, increasing the number of terms in the Fourier series from 2 to 4 doubles the best period of the LS periodogram, to 2.1 h. Therefore, it is possible that the in-phase lightcurve is more complex than a bimodal one, with four maxima and four minima. The FALC algorithm also yields the same results. Considering that the absolute magnitude provided by the MPC is about +19.9 mag, a size between 300 and 600 meters can be estimated, so this asteroid is a relatively large object well below the cohesionless spin barrier.

\subsection{(333284) 1999~PJ$_{1}$}
\label{sec:1999PJ1}
The NEA-Amor 1999~PJ$_{1}$ was observed on the evening of August 22, 2025, with the BFOSC. The sky was clear, with a few passing clouds, and the Moon was absent. The asteroid was moving at about 4 arcsec/minute across the celestial sphere, and 30-second exposures were taken with $BVR_cI_c$ filters for taxonomic classification. For calibration, the Landolt field PG1633 with a very similar airmass on the horizon was used. The rotation period was already known with good precision (about 6.20 h), so lightcurve photometry was not performed. The colour indices of 1999~PJ$_{1}$ are the following: $B-V= 1.19 \pm 0.09 $, $V-R_c= 0.44 \pm 0.08 $ and $V-I_c= 0.60 \pm 0.06$ mag. The color indices are somewhat uncertain due to the asteroid's high apparent $V$ magnitude of about 17.4 and its rapid motion on the sky, which limit the SNR. Based on the reflectance values, the asteroid appears consistent with a V-type in 81\% of cases. However, the index $B-V = 1.19 \pm 0.09$ mag is anomalously red for a basaltic V-type, which typically has a value near 0.81 \citep{Dandy2003}, so the V match appears to be driven almost entirely by the low $V-I_c = 0.60 \pm 0.06$. For this reason, we consider the attribution to V-type uncertain. The $H_V$ magnitude from calibrated $V$ value is $18.4 \pm 0.2$ mag, and the estimated diameter for the taxonomic class is $D=0.5 \pm 0.1$ km.

\subsection{(152664) 1998~FW$_{4}$}
\label{sec:1998FW4}
The NEA-PHA 1998~FW$_{4}$ was observed with BFOSC on the night of September 18, 2025. The session was characterized by clear skies and the absence of the Moon. We take four $BVR_cI_c$ filtered images with a 30-second exposure. For calibration, we use the Landolt fields PG2336 and PG2213, both of which have approximately the same airmass as the target. The magnitude values for the asteroid obtained in the different filters are the mean of those from calibration using the two Landolt fields. For this asteroid, the rotation period was already known; therefore, we did not collect a lightcurve. We found the following values: $B-V = 0.75 \pm 0.03$, $V-R_c = 0.44 \pm 0.03$, and $V-I_c = 0.77 \pm 0.14$ mag. The last color index has a high uncertainty due to the large systematic variation in the asteroid's $I_c$ magnitude during the 40-minute observation period (interspersed with imaging of the two Landolt fields), which does not involve the $B$ and $V$ bands and, marginally, the $R_c$ band. This could be due to macroscopic color variations on the asteroid's surface, visible only at longer wavelengths.\\
Using the reflectance curve and the Monte Carlo probability values, the asteroid belongs to the X-complex. The $H_V$ magnitude from calibrated $V$ value is $19.7 \pm 0.3$ mag, and the estimated diameter for the taxonomic class is $D=0.5 \pm 0.2$ km.

\subsection{2025~FA$_{22}$}
\label{sec:2025FA22}
\begin{figure}
    \centering
    \includegraphics[width=0.8\textwidth]{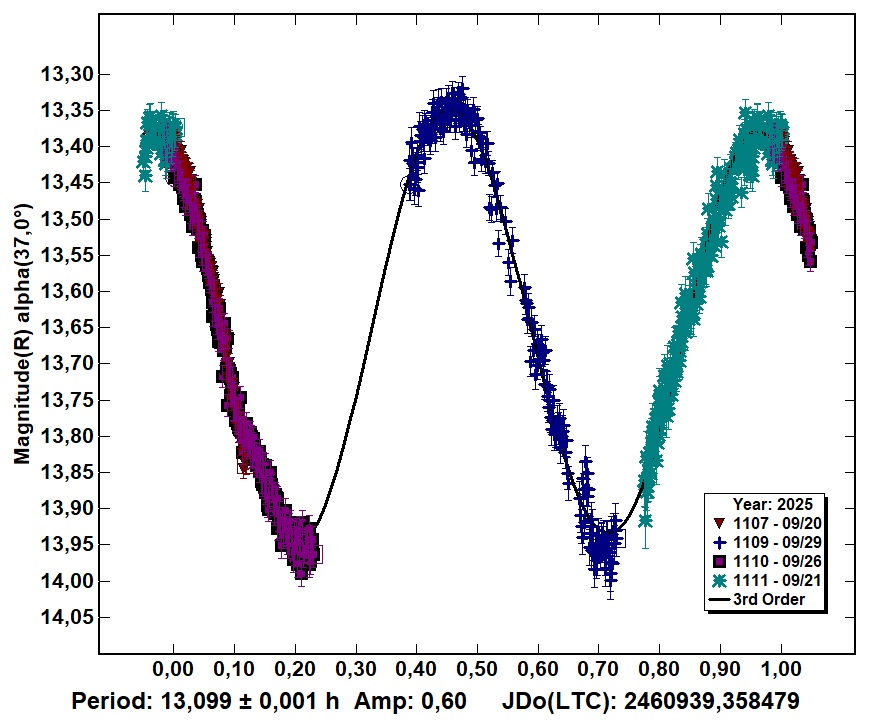}
    \caption{The phased lightcurve of NEA 2025~FA$_{22}$ obtained with multiple sessions. The best rotation period is about 13.1 h, with an amplitude of about 0.6 mag.}
    \label{fig:2025FA22}
\end{figure}

The Apollo-PHA asteroid 2025~FA$_{22}$ on September 18, 2025, made a close flyby with the Earth at a distance twice the lunar distance. This approach was used for an IAWN (International Asteroid Warning Network) campaign to characterize the asteroid from both orbital and physical perspectives. So, 2025~FA$_{22}$ was initially observed on the evening of Sep 18, using the TANDEM system to perform astrometry of several tracks left by the asteroid in the FoV, which was moving across the sky at about 100 arcsec/minute. Some observations were submitted to the MPC and accepted for publication \citep{Palmiotto2026}. On the nights of September 20 and 21, two photometric sessions of approximately 3 hours each were conducted using the telescope of the Virgil Observatory with the $R_c$ filter, in order to determine the rotation period. However, this proved longer than expected, so photometric observations continued with BFOSC on the nights of September 26 and 29. With all these sessions, the phased lightcurve appears almost complete. The best-fitting lightcurve shows an amplitude of 0.6 mag and a period of approximately 13 hours, see Fig.~\ref{fig:2025FA22}. The photometric data for the lightcurve were also shared with the IAWN team, who, using all available data, confirmed the period value. \\
On the clear and very transparent night of September 26, several images were also taken with the $BVR_cI_c$ filters and the PG2213 Landolt field at the same airmass to obtain color indices. The colors of the asteroid are $B-V=0.66 \pm 0.05$, $V-R_c=0.46 \pm 0.03$, and $V-I_c=0.86 \pm 0.04$ mag, and from the reflectance, the asteroid is T-type with a probability of 36.2\% and X-complex with a probability of 34\%. Instead, when using color indices, the asteroid appears to be an X-complex object \citep{Dandy2003}. Given the small difference in reflectance between X and T (see Fig.~\ref{fig:reflectance} and Fig 4 of \cite{DeMeo2013}), the near-equal probability appears justified, and we adopt the X-complex as a more probable classification. The $H_V$ magnitude from calibrated $V$ value is $21.4 \pm 0.3$ mag, and the estimated diameter for the taxonomic class is $D=0.2 \pm 0.1$ km.

\subsection{(138205) 2000~EZ$_{148}$}
\label{sec:2000EZ148}
\begin{figure}
    \centering
    \includegraphics[width=0.8\textwidth]{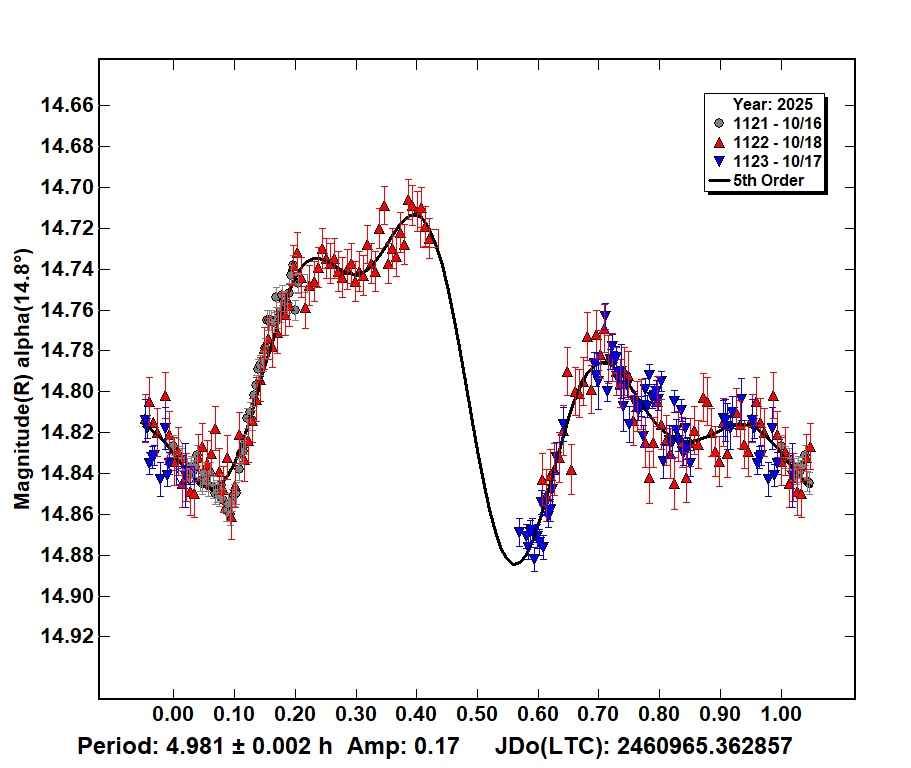}
    \caption{The phased lightcurve of NEA 2000~EZ$_{148}$ was obtained with three different sessions, for a total of about 7.2 hours of observation time. The best rotation period is consistent with about 4.98 h, with an amplitude of 0.17 mag.}
    \label{fig:2000_EZ148}
\end{figure}

The Apollo asteroid 2000~EZ$_{148}$ was first observed with BFOSC on the clear night of October 16, 2025. During this first session, a set of three $BVR_cI_c$-filtered images was taken, together with the Landolt field PG2213 as a calibration, at the same airmass. Next, a first photometric run lasting about 1 hour with the $R_c$ filter was conducted. The asteroid has an absolute magnitude of +15.04 mag in the JPL Small-Body Database, and, given its large size, we expected a long rotation period. Instead, from this short lightcurve, a large amplitude of approximately 0.1 mag is observed. The next night, a longer 2.2-hour photometric session was conducted, but the sky was partially cloudy, and the data are of only average quality. We conducted another four-hour-long session on the clear night of October 18, using the Virgil Observatory telescope in the $R_c$ band. Putting all together, with the FALC algorithm, we found the best rotation period of about 4.98 h with an amplitude of 0.17 mag, see Fig.~\ref{fig:2000_EZ148}.\\
The color indices from the BFOSC data taken on October 16 are the following: $B-V=0.84 \pm 0.03$, $V-R_c=0.503 \pm 0.003$, and $V-I_c=0.91 \pm 0.02$ mag, and from the reflectance, the asteroid belongs to the S-complex. This supports the uncertain taxonomic classification from \cite{Binzel2004}. From this classification, an absolute magnitude of $15.5 \pm 0.2$ mag and an effective diameter of $2.4 \pm 0.5$ km are obtained. 

\subsection{(843079) 2016 CB$_{32}$}
\label{sec:2016CB32}
This highly eccentric Amor was observed in a single session on the morning of Nov 29, 2025, with BFOSC before an astrometric session on the interstellar comet 3I/ATLAS \footnote{\url{https://minorplanetcenter.net/mpec/K25/K25N12.html}}. The sky was clear and without the Moon, with a FWHM of about 2.2 arcsec. We have taken four series of $BVR_cI_c$ images, and as calibration, the Landolt field RU 149 at the same airmass and in the same part of the sky. Next, we obtain a dense photometric series with the $R_c$ filter, which is approximately 35 minutes long. Unfortunately, in the second part of the session, the asteroid passed prospectively very close to two field stars, and several photometric points were eliminated to avoid altering the lightcurve. The rotation period for this asteroid can not be computed; we can only say that the amplitude is greater than 0.04 mag. It will be difficult to estimate the period based on other observations, because this asteroid reached its closest approach to Earth on November 19 at a geocentric distance of 0.197 au, and that same distance will not be reached for decades.\\
The color indices were the following: $B-V=0.72 \pm 0.03$, $V-R_c=0.46 \pm 0.02$, and $V-I_c=0.77 \pm 0.02$ mag, and from the reflectance and Monte Carlo probabilities, the asteroid has a best match with the X-complex. From this classification, an absolute magnitude of $18.9 \pm 0.3$ mag and an effective diameter of $0.7 \pm 0.3$ km are obtained. 

\begin{figure}
    \centering
    \includegraphics[width=0.8\textwidth]{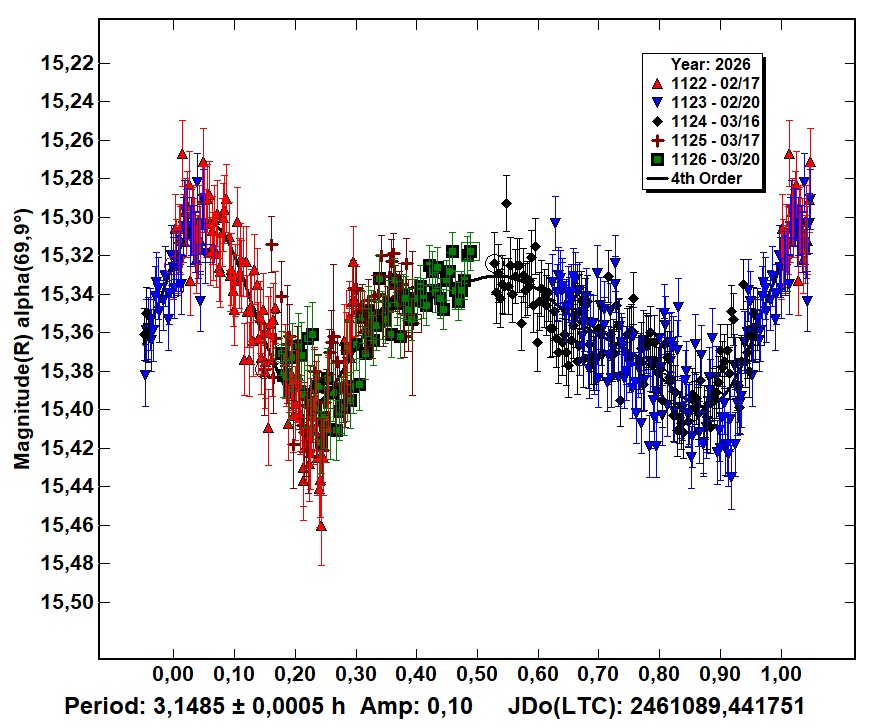}
    \caption{The phased lightcurve of NEA 1992~SY obtained with 4 sessions 5 hour long. The best rotation period is about 3.15 h, with an amplitude of about 0.10 mag.}
    \label{fig:1992_SY}
\end{figure}

\subsection{(52340) 1992 SY}
\label{sec:1992SY}
The Apollo asteroid 1992 SY was observed on the night of February 17, 2026. The sky was clear with stable transparency and without wind. The apparent magnitude of the target was about +16.3, with an angular speed of 5.53 arcsec/min. We observed it with the BFOSC in bin 2 mode, with a maximum exposure time of 20 s to obtain a star-like image on the CCD. We take 10 images for each $BVR_cI_c$ filter and use the Landolt field PG0918 at a similar airmass to calibrate the magnitudes. The colour indices result the following: $B-R_c = 1.29 \pm 0.08$,  $B-V = 0.87 \pm 0.08$, $V-R_c = 0.43 \pm 0.02$ and $V-I_c = 0.71 \pm 0.04$ mag. With these values, the reflectance spectra indicate a V-type asteroid with a probability of 45 percent. Given the low probability and possible strong reddening effect at high phase angle of $\alpha=70$~deg, our classification of 1992 SY as the V-type remains uncertain. Assuming these taxonomy classes, the absolute magnitude is $18.3 \pm 0.2$ mag, and the corresponding diameter is $D=0.5 \pm 0.1$.\\
Regarding the rotation period, we perform dense photometry with the $R_c$ filter for about 1 hour, and we repeat it on February 20 and March 16 and 20 for another 4 hours. The phased lightcurve from the FALC algorithm shows an amplitude of about 0.10 mag and a best-fit rotation period of about 3.15 hours, see Fig.~\ref{fig:1992_SY}.

\end{document}